# Dielectrics for Two-Dimensional Transition Metal Dichalcogenide Applications

Chit Siong Lau,[1, *] Sarthak Das,[1] Ivan A. Verzhbitskiy,[1] Ding Huang,[1] Yiyu Zhang,[1] Teymour Talha-Dean,[1, 2] Yiyu Zhang,[1] Wei Fu,[1] Dasari Venkatakrishnarao,[1] and Kuan Eng Johnson Goh[1, 3, 4]

[1]Institute of Materials Research and Engineering (IMRE), Agency for Science,
Technology and Research (A*STAR), 2 Fusionopolis Way,
Innovis #08-03, Singapore 138634, Republic of Singapore
[2]Department of Physics and Astronomy, Queen Mary University of London, London, E1 4NS, United Kingdom
[3]Department of Physics, National University of Singapore, 2 Science Drive 3, 117551, Singapore
[4]Division of Physics and Applied Physics, School of Physical and Mathematical Sciences,
Nanyang Technological University, 50 Nanyang Avenue 639798, Singapore

Despite over a decade of intense research efforts, the full potential of two-dimensional transition metal dichalcogenides continues to be limited by major challenges. The lack of compatible and scalable dielectric materials and integration techniques restrict device performances and their commercial applications. Conventional dielectric integration techniques for bulk semiconductors are difficult to adapt for atomically thin two-dimensional materials. This review provides a brief introduction into various common and emerging dielectric synthesis and integration techniques and discusses their applicability for 2D transition metal dichalcogenides. Dielectric integration for various applications is reviewed in subsequent sections including nanoelectronics, optoelectronics, flexible electronics, valleytronics, biosensing, quantum information processing, and quantum sensing. For each application, we introduce basic device working principles, discuss the specific dielectric requirements, review current progress, present key challenges, and offer insights into future prospects and opportunities.

## I. INTRODUCTION

Two-dimensional (2D) semiconducting transition metal dichalcogenides (TMDs) are a class of materials with the general formula $MX_2$, where M is a transition metal (e.g., Mo and W) and X is a chalcogen (e.g., S, Se or Te).[1–6] Their properties such as atomically thin geometries, pristine surfaces lacking dangling bonds, layer-dependent direct band gaps, and spin-valley coupling have generated intense interest in both academia and industry for next-generation applications in many diverse fields.

Despite tremendous progress since their isolation, 2D TMD based electronic devices are still unable to successfully make the 'lab-to-fab' transition. Initial research efforts were heavily focused towards the study of the different 2D TMD semiconducting materials. However, in many cases, developments are limited due to a lack of compatible dielectrics, scalable growth, processing, and device engineering techniques.[7–10] Conventional complementary-metal-oxide-semiconductor (CMOS) techniques are typically not easily adaptable from 3D bulk materials while preserving interface quality;[9, 11] and without altering or damaging 2D materials that can be more delicate due to their atomically thin nature. For example, ALD grown $HfO_2$ is a commercially viable dielectric used in Intel's 14 nm node,[12] but while ALD $HfO_2$ has also been reported for 2D $MoS_2$ transistors,[13] the device performance suffers from poor dielectric growth. The lack of compatible dielectrics and integration techniques represent key bottlenecks for the field of 2D TMDs. Furthermore, dielectric requirements can differ greatly between applications. For example, a field-effect transistor for logic operations requires thin dielectrics with high dielectric constants for fast operation and energy efficiency; a phototransistor requires dielectrics transparent over the photon wavelengths of interest; flexible electronics require high mechanical flexibility and stability; quantum information processing desires dielectrics with low charge and magnetic noise. There is yet no commercially viable technique that can *concurrently* fulfill all the various dielectric requirements for specific applications.

In this review, we start with a brief introduction to various dielectric synthesis techniques that have been demonstrated or are promising for integration with 2D TMDs. For each technique, we describe the fundamental growth principles and associated advantages and limitations for 2D TMD integration. Next, we review 2D TMD dielectric integration for various applications including nano-electronics, optoelectronics, flexible electronics, quantum information processing, quantum sensing, biosensing, valleytronics, and spintronics. In each section, we first introduce the basic working principles of the relevant TMD devices, before highlighting the specific requirements of dielectrics for that application. We then review the progress made before concluding with a discussion on key challenges and future outlook for each application.

* aaron_lau@imre.a-star.edu.sg

## II.  DIELECTRIC SYNTHESIS

In this section, we introduce common and emerging dielectric synthesis techniques and discuss their applicability for 2D TMD devices, such as process-induced impact on the quality of 2D TMD channel and interface, potential scalability, and challenges.

### A.  Mechanical and chemical exfoliation

Van der Waals (vdW) materials consist of individual atomic planes with strong in-plane bonding but weak out-of-plane vdW bonding. High-quality ultrathin vdW dielectric layers can be mechanically exfoliated using adhesive tapes from their corresponding bulk layered materials, which are usually grown by chemical vapor transport (CVT) (Figure 1a).[7, 14–21] For chemical exfoliation in solution, these involve assistance by sonication, ion intercalation and electrochemical methods.[22, 23] Insulators like hexagonal boron nitride (hBN) flakes are commonly synthesized from bulk growth followed by subsequent exfoliation and are among the most widely used vdW dielectric layers in 2D material-based devices due to their high quality.

Mechanical exfoliation of bulk single crystals can produce high-quality, single-crystalline 2D films of mono- and few-layered hBN (Figure 1a). Subsequent transfer enables the stacking of dielectric layers on 2D materials.[24, 25] The high quality of the dielectric and vdW interface makes mechanical exfoliation an ideal tool for fundamental studies in the laboratory. However, it is not suitable for the large-scale production of 2D materials, as exfoliated crystals are typically random in size (~5 –50 μm), shape, number of layers, and position. While solution processed chemical exfoliation offers potential scalability of monolayer or few-layer vdW dielectrics, the current limitation remains the broad distribution of layer thickness and lateral size, low yield, low quality and high tendency of re-stacking.[22, 23, 26]

### B.  Evaporation and sputtering

Evaporation and sputtering are commonly used, scalable techniques to prepare various dielectrics on bulk (Figure 1b).[27–29] In sputtering, the dielectric material source target is bombarded with Argon or Nitrogen ions which erupts the dielectric material, while in evaporation, dielectric materials are deposited by heating the source material with thermal heating or an electron beam. The substrates for deposition are placed in front of the targets at an appropriate distance. Sputtering typically requires the use of harsh environments, such as plasma, which can easily damage 2D materials due to their atomic-scale thickness.[30, 31] The high energies of the erupted dielectric materials in both sputtering and evaporation can also lead to damage at the dielectric/TMD interface that deteriorates the electrical properties of 2D TMD devices.[32, 33]

### C.  Native oxidation

Inspired by the industrial fabrication of $SiO_2$ dielectric on Si,[34] the partial oxidation of 2D TMDs can transform the topmost layers into their native oxides while leaving the bottom layers intact (Figure 1c).[35–42] The process may lead to atomically abrupt and defect-free interfaces as was demonstrated with vdW semiconductor $Bi_2SeO_2$ partially oxidized to insulating $Bi_2SeO_5$ at elevated temperatures (Figure 1c).[41, 42] $Bi_2SeO_2$ can be converted into $Bi_2SeO_5$ in a layer-by-layer manner, forming an atomically sharp interface. However, this native oxidation method is highly limited to the types and thicknesses of 2D semiconductors and is challenging for wide applications. Achieving such precise control may also be challenging when fabricating monolayer devices.

### D.  Chemical vapor deposition

2D synthetic hBN dielectrics can also be grown by chemical vapour deposition (Figure 1d).[21, 43–47] The synthesis of hBN is carried out at high temperatures under a hydrogen and/or nitrogen atmosphere using Cu foil as catalyst and borazine as a precursor. The growth rate can be controlled by the borazine bubbling rate. However, it is challenging to grow multilayer hBN with good uniformity on a large scale. Furthermore, the subsequent transfer process of hBN layers onto 2D TMDs can introduce impurities, cracks, wrinkles, tears, or trapped gas bubbles resulting in poor interface quality.[48]

### E.  Atomic layer deposition

ALD is a thin-film deposition method based on surface chemical reactions (Figure 1e) .[49] Due to the mechanisms of surface self-saturated reactions, ALD has advantages such as film thickness control at the atomic scale and excellent conformality. Generally, one ALD cycle is composed of four steps: (i) precursor exposure, (ii) purging, (iii) counter-reactant exposure, and (iv) purging (Figure 1e). Because of the sequential exposure of the precursor and counter reactant separated by the purging steps, chemical reactions only occur on the surface and strongly depend on the surface properties. Compared with sputtering, ALD is performed under milder conditions. However, as the surfaces of most 2D materials are inert without dangling bonds, the precursors are only physically adsorbed on 2D materials which can lead to non-uniform growth. Various approaches have been developed to improve the uniformity of ALD-grown dielectrics, such as creating defects 2D material surfaces through plasma or wet etching[50–54] as well as deposition of metallic or molecular seeding layers[55–57] (Figure 1d). Unfortunately, the deposition of seeding layers can compromise the TMD surface quality, while discrete molecules can make homogeneous film growth challenging.[58]

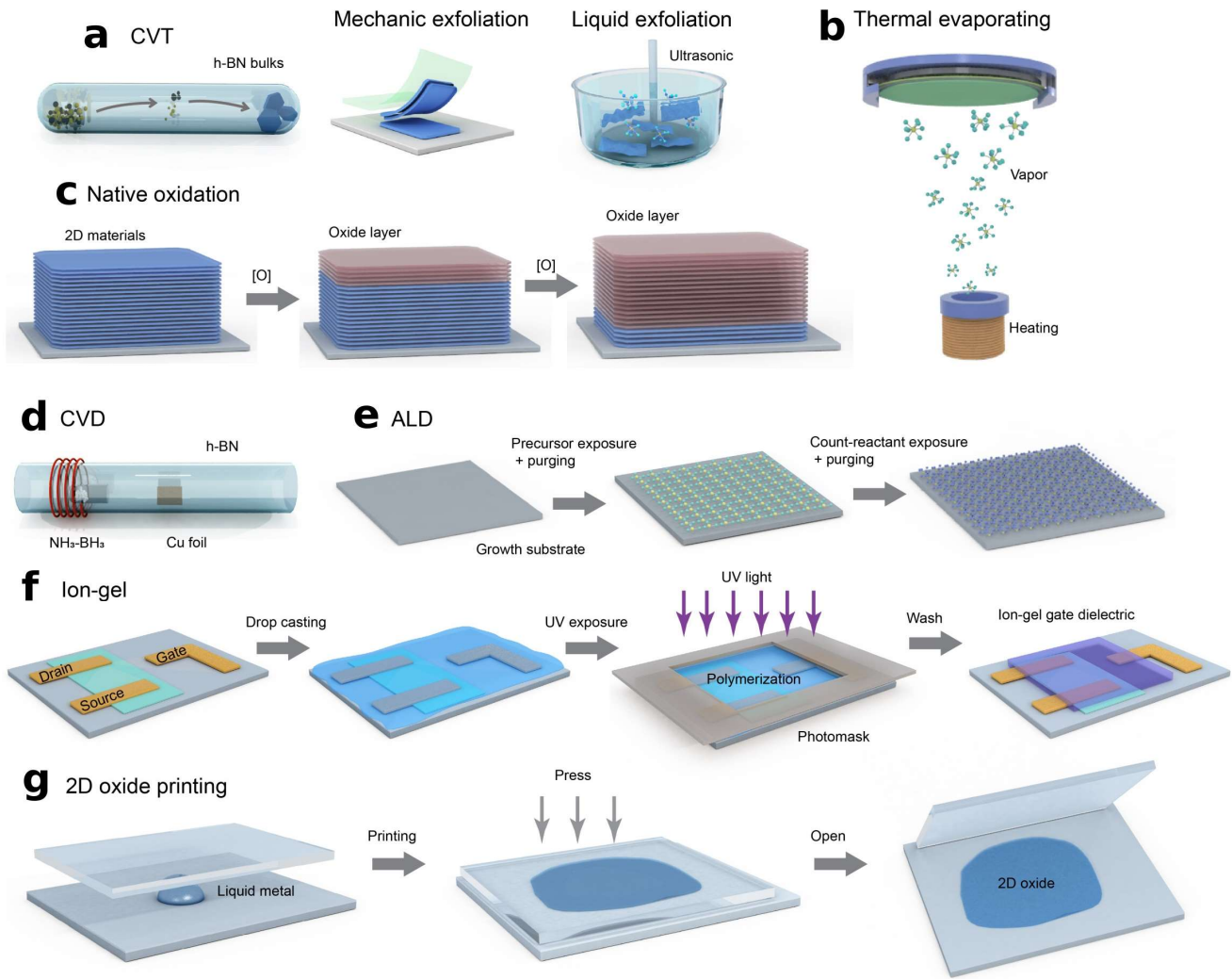

FIG. 1. Dielectric synthesis techniques. (a) Chemical vapour transport (CVT) growth of bulk dielectric materials. These bulk materials are then mechanically exfoliated with an adhesive tape or liquid exfoliated in solution. (b) Evaporation/sputtering involves the transfer of atoms from a bulk source material to the target substrate. (c) Native oxidation. (d) Chemical vapour deposition (CVD) growth of hexagonal boron nitride. (e) Atomic layer deposition (ALD). (f) Ion-gel via drop casting and subsequent UV photolithography. (f) 2D oxide printing from liquid metals.

### F. Molecular beam epitaxy

Molecular beam epitaxy (MBE) is a widely used technique in semiconductor fabrication for the controlled epitaxial growth of thin films in an ultra-high vacuum environment (Figure 1f). The precursors are separately heated in effusion cells until sublimation, after which the gaseous elements condense and react on the deposition substrate. MBE has been used to grow high quality epitaxial single crystal dielectrics such as $CaF_2$[59–62] and $SrTiO_3$[63, 64] that can form quasi vdW interfaces with 2D TMDs and serve as back gate dielectrics. While high quality films can be produced, epitaxial growth typically require specially prepared substrates, and its direct, pinhole-free growth on 2D materials remain challenging.

### G. Ion-gel

Ion gels are composite electrolyte materials composed of polymer networks integrated with an ionic liquids and combine the excellent electrical properties of ionic liquids and the flexible mechanical properties of polymers (Figure 1g).[65–67] They are ionic conductors but electronically insulating dielectrics that can form electric double layers from the accumulation of space charges at interfaces.[68, 69] A key advantage is their high specific capacitance (typically > 1 $\mu$Fcm$^{-2}$) enabling energy efficient device operation at low voltages. Low temperature, large-area solution processing is also possible, and their compatibility with flexible substrates is useful for flexible electronic applications. The long-range polarizability of ion-gel electrolytes also allows for fabrication ease where gate electrode alignment requirements are less stringent. Ion-gels can be integrated with devices through low-cost and high throughput techniques such as aerosol and inkjet printing, spray coating, and spincoating. These techniques are potentially compatible with large-scale manufacturing, but may lead to possible solvent contamination of semiconductor channels and nonuniformity. Large faradaic leakage currents are also a challenge.

### H. Liquid metal exfoliation

Liquid metal (LM) exfoliation is an emerging strategy to produce ultrathin metal oxide dielectrics in a potentially scalable and low-cost fashion (Figure 1h).[70–72] LMs are metals and alloys whose melting points are close to or below room temperature, e.g., gallium, EGaIn (gallium, indium) and gallinstan (gallium, indium, tin).[73–75] Many LMs can rapidly form a self-limiting, ultrathin ( 1-5 nm) oxide skin on their surfaces when exposed to an oxygen-containing atmosphere.[71, 76] The oxide formation kinetics and film properties are dependent on environmental and surface conditions, e.g., oxygen partial pressure, temperature, and surface tension. Strategies involving environmental control and surface functionalisation may therefore lead to tuneable oxide properties to engineer specific device functionalities for targeted applications.[77]

More interestingly, preferential oxide formation on LM surfaces can be predicted through thermodynamic considerations.[70, 71, 78] For example, in galinstan (Ga, In, Sn), the surface oxide is exclusively composed of gallium oxide. Several other oxides such as HfO$_2$,[70] Al$_2$O$_3$,[70] and Gd$_2$O$_3$[70] have been demonstrated (Table 1). These ultrathin LM oxides can be exfoliated and transferred to other substrates in many ways including touch or squeeze printing,[70, 79] blade coating and roll transfer,[80, 81] or liquid phase exfoliation through sonication or bubbling.[70] The transferred oxide films are shown to be uniform, large area ( cm sizes), and ultrathin (<5 nm) with high consistency. However, limitations exist. For example, the degree of film crystallinity is typically low. Residual LM nanoparticles can also be difficult to remove and may present problems when integrating with 2D TMDs for device applications. So far, there are limited reports on its integration with 2D TMDs for device applications. Future research should develop material processing techniques and film tuning strategies to address various applications.

## III. APPLICATIONS

### A. Nanoelectronics

Field-effect transistors (FETs) for 2D TMDs are promising candidates for 'More than Moore' devices (Figure 3a).[3, 10, 82–86] In most nanoelectronic devices such as the FET, the dielectric is an electrically insulating layer separating the carriers in the channel, i.e., the 2D semiconducting TMD, from the conductive gate electrode that modulates the overall device current. The relationship between dielectric properties and overall device performance is complex.[7, 8, 49] Nevertheless, we attempt to summarize the gate dielectric requirements into three categories: quality, scalability, and compatibility (Figure 2). For each category, we discuss specific dielectric properties and assess their impact on electronic devices (Figure 2b). We then review dielectric integration approaches based on amorphous oxides, crystalline dielectrics, and native oxides.

#### 1. Scalability

The downsizing of transistors for denser circuits with better performance and greater energy efficiency must preserve gate electrostatics. This requires a high capacitance $C$. Specifically, $C$ >3 $\mu$Fcm$^{-2}$ for a carrier density $10^{13}$cm$^{-2}$ at gate voltage $V_{GS}$ <1 V.[8, 83] As $C \propto \frac{\kappa}{t}$, where $\kappa$ is the dielectric permittivity and $t$ the dielectric thickness, ideal materials should have high $\kappa$, e.g., HfO$_2$ ($\kappa_{HfO_2}$ =23, $\kappa_{SiO_2}$ =3.9), [18] or $t$ should be easily decreased. This scaling is typically expressed via the equivalent oxide thickness (EOT) which describes the thickness of SiO$_2$ that induces the same capacitance as the dielectric used. EOT of a dielectric A is defined as EOT$_A$=$t_A$($\frac{\kappa_{SiO_2}}{\kappa_A}$) and is useful to benchmark different dielectrics against the industry standard SiO$_2$. The International Roadmap for Devices and Systems 2020 update sets key transistor performance targets such as an EOT of 0.6 nm. However, the physical reduction of $t$ must also delicately balance the corresponding increase in gate leakage currents, which should remain below 0.01 Acm$^{-2}$ at $V_G$ <0.7 V (low power applications) and 1 Acm$^{-2}$ (high power applications). This consideration thus favours high-$\kappa$ dielectrics that can provide high capacitances while maintaining a larger $t$ for reduced gate current leakage. Another requirement to maintain device integrity and reliability is a breakdown field 10 MVcm$^{-1}$.

These scalability requirements are meant to enhance transistor performance. Small EOTs lead to devices with efficient gate control and small subthreshold swings, i.e. how quickly a transistor turns on and off, that can approach the room temperature thermionic limit of 60 mV/dec. Resulting high current on/off ratios and low operating voltages increase energy efficiency while high breakdowns fields contribute towards reliable operation over a typical lifetime of 10 years.[7, 8, 59, 88]

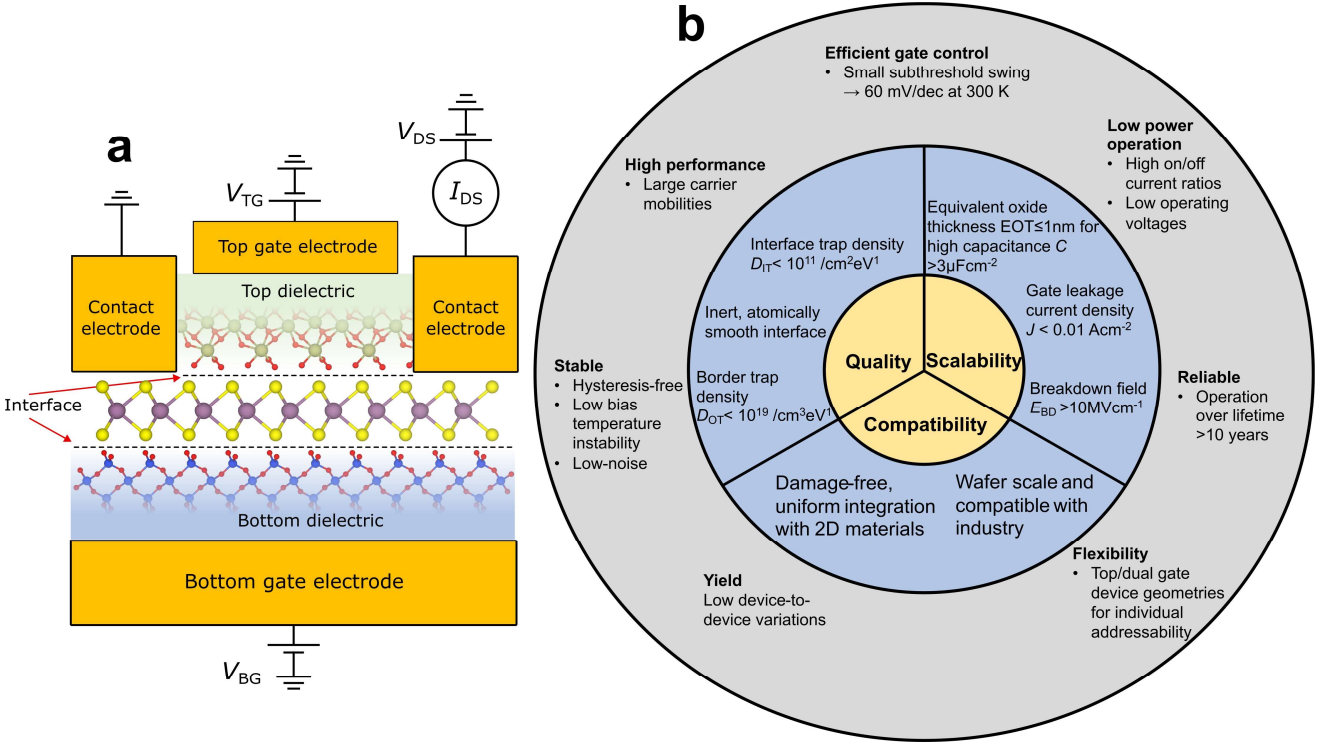

FIG. 2. (a) Schematic of a typical 2D semiconducting TMD FET. Devices are frequently fabricated on a doped Si substrate with a $SiO_2$ layer that can also operate as a back gate by applying a voltage $V_{BG}$. A current $I_{DS}$ is driven across the device with a bias voltage $V_{DS}$ across the contacts, that can also be modulated by a top (bottom) gate voltage $V_{TG}$ ($V_{BG}$). (b) Schematic overview of the requirements for dielectrics to realise 2D TMD-based FETs.[8, 83, 87]

### 2. Quality

Both the dielectric quality and the dielectric/semiconductor interface quality are critical for device performance. Ideally, we require an inert and atomically smooth interface with a low interface trap density $D_{IT} < 10^{10}$ cm$^{-2}$eV$^{-1}$.[8, 83, 89] This reduces carrier scattering from interface roughness and charged impurities and retains the high carrier mobility of the 2D TMD active channel.[90] High-$\kappa$ dielectrics can screen charged impurities to enhance carrier mobility, but this may come with increased remote phonon scattering contribution from the dielectric.[91–93] Remote phonon scattering is generally strongest for high-$\kappa$ dielectrics that allow polar vibrational modes, and weakest in dielectrics with small $\kappa$.[89, 92]

Besides carrier mobility, $D_{IT}$ is related to the sub-threshold swing (SS) as SS=$2.3\frac{k_BT}{q}(1 + \frac{qD_{IT}}{C_D})$, where $q$ is the electron charge, $k_B$ the Boltzmann constant, $T$ the temperature, and $C_D$ the dielectric capacitance. For SS to approach the thermionic limit, $\frac{qD_{IT}}{C_D} \propto D_{IT}\text{EOT}_D \to 0$. Therefore, efficient gate control requires low $D_{IT}$ in addition to small EOTs. $D_{IT}$ is not only dependent on the dielectric, but also the electrically active defect density of the 2D semiconducting channel material. This can vary significantly based on the synthesis process, e.g., exfoliated flakes are lower in defect densities ($\sim 10^{11}$ cm$^{-2}$) compared to CVD grown flakes ($\sim 10^{13}$ cm$^{-2}$).[94]

Border traps are located within the gate dielectric at a few nanometers away from the interface and can capture carriers from the 2D semiconductor via tunnelling.[95] The border trap defect density $D_{OT}$ should ideally be $< 10^{19}$ cm$^{-3}$eV$^{-1}$[96] to minimise charge trapping events that increase hysteresis, noise, and bias temperature instability (BTI) that degrade device stability and reliability. BTI causes shifts in device parameters, e.g., threshold voltage, that accumulate over time until the operating point changes lead to device failure. Another consideration for device stability and reliability is the fraction of electrically active defects, determined by the relative alignment of the dielectric defect band to the 2D semiconductor band edges (Figure 3a).

### 3. Compatibility

To realise the commercial potential of 2D semiconductor devices, dielectric integration must be compatible with industrial semiconductor fabrication. Top-gated devices are standard structures for integrated circuits, while very-large-scale-integration (VLSI) require dual-gate structures.[83] Suitable gate dielectrics should therefore allows uniform, damage-free, top-down deposition onto 2D semiconductors with high yield and homogeneity on the wafer scale with sub-nanometre thickness control. Otherwise, automated large-area transfer techniques must be developed.[25] These processes should also be compatible with front-end-of-line (FEOL) or back-end-of-line (BEOL) applications. FEOL is the part of integrated circuit applications where components such as transistors, capacitors, and resistors are fabricated and process temperatures cannot exceed 700 °C.

BEOL are for interconnects and diffusion barriers and cannot withstand temperatures exceeding 400 °C.[10] For heterogenous integration, dielectrics with potential for selective etching are important for opening holes/vias to pattern BEOL contacts to buried 2D semiconductors.[84]

These are non-trivial requirements. Due to a lack of dangling bonds on many 2D TMD surfaces, they are chemically inert and inhibit dielectric precursor nucleation. The atomically thin geometries of 2D semiconductors are susceptible to damage from deposition, transfer, pre-treatment, and selective etching processes. Processing can leave unwanted residue or water and other gaseous adsorbates from an ambient environment. The resulting surface contamination affects dielectric integration and cause unintentional doping and degradation of the 2D FET.[55]

Having introduced the dielectric requirements based broadly on scalability, quality, and compatibility, we next review dielectric integration approaches based on amorphous oxides, crystalline dielectrics, and native oxides.

### 4. Amorphous oxides

Amorphous oxides are typically deposited using conventional semiconductor techniques such as direct physical vapour deposition or ALD.[49] Several oxides have been explored for 2D TMD devices including $SiO_2$,[97] $HfO_2$,[13, 52, 53, 55–57, 90, 98–100] $Al_2O_3$,[58, 101] $TiO_2$,[101, 102] $Er_2O_3$,[103] and $ZrO_2$.[104, 105] While ALD offers uniform, conformal, high-$\kappa$ dielectric deposition with sub-nanometre thickness control on the wafer scale, the main challenge is the inert surfaces of 2D TMDs that inhibit precursor nucleation, leading to poor initial growth and rough interfaces (Figure 3b, c).

Efforts to circumvent this can be broadly classified into (1) surface treatments of the 2D TMD surface,[52–54] and (2) using seed layers. Surface treatments include UV-$O_3$, $O_2$ plasma, and $Ar^+$ ion to generate reactive sites through limited defect or adsorbent introduction but can degrade or alter the 2D TMD properties (Figure 3d, e). While this may be acceptable in few-layer films, the surface damage is catastrophic for monolayer device performance. Seed layers can support precursor nucleation but must themselves be deposited with minimal damage to the 2D semiconductor, e.g., Al and Hf[56] seed layers promote uniform oxide growth but metal evaporation can damage the 2D semiconductor surface. Organic molecular seed layers have been explored but most are low-$\kappa$ and the discrete nature of molecules can lead to large variability and poor uniformity.[58] However, a more promising approach uses PTCDA molecular crystal seeding where the thickness can be controlled down to 0.3 nm through self-limited epitaxy.[57] Devices achieved 1 nm EOT, though the effective $\kappa$ was limited when scaled to small EOT.

A challenge for amorphous oxides is the higher defect densities. A potential solution is to select dielectrics with defect bands energetically located far away from the Fermi level of 2D TMDs as charge trapping is highly sensitive to this alignment (Figure 3a).[106] While such defects are problematic for transistors in conventional von-Neumann architectures, an interesting recent approach is the use of 2D materials for non-von-Neumann applications where the influence of such defects is less important, or possibly even beneficial. For example, a key bottleneck in von-Neumann computing is data shuttling from the physical separation of memory and logic elements, leading to area and energy inefficient hardware. 2D memtransistors that offer analog and non-volatile memory can potentially circumvent this bottleneck in non-von-Neumann computing.[107–110] Such analog non-volatile memory can be achieved by exploiting otherwise detrimental charge trapping at or near the dielectric/semiconductor interface,[110] allowing the demonstration of 2D TMD based neuromorphic and biomimetic applications. For example, bio-inspired devices have been demonstrated using in a split-gated $MoS_2$ device with hydrogen silsesquioxane (HSQ) gate dielectric that can mimic the auditory cortex of a barn owl [111] or be used for energy efficient and bio-realistic spiking neural networks [112, 113]. ALD grown 50 nm $Al_2O_3$ on Pt/TiN/p$^{++}$ Si stack with 2D TMDs were used to demonstrate simulated annealing for Ising type spin systems,[114] biomimetic collision detection based on the escape response in locusts,[115] monolithically integrated, multipixel, bioinspired neural networks,[116] bioinspired machine vision,[117] and insect-inspired collision detectors [118]. Similarly, inherent stochastic charge trapping and detrapping in gate stacks that lead to cycle-to-cycle variation in device characteristics have been exploited for applications in stochastic computing [119, 120] and hardware security [121, 122]. Similar architectures based on ALD grown 50 nm $Al_2O_3$/Pt/TiN stack for $MoS_2$ memtransistor exploited the programmable stochasticity to demonstrate logic gate based arithmetic operations,[123] energy efficient hardware implementation of a Bayesian network,[124] and probabilistic synapses and reconfigurable neurons for Bayesian neural networks [125]. Hardware security applications can also utilize this architecture, with demonstrations of logic locking,[126] physical unclonable functions,[127] anticounterfeit measures,[127] optical watermarks,[127] camouflaging, [127] and true random number generators [128].

### 5. Crystalline dielectrics

Crystalline dielectrics that form inert interfaces with 2D semiconductors are theoretically predicted to have the lowest defect densities with narrower defect bands compared to amorphous oxides. The most widely used and studied dielectric for 2D materials is crystalline hBN.[5, 18–20, 44, 46, 94, 130, 131] As a dielectric for graphene, hBN enabled many breakthrough observations of physics such as unconventional superconductivity and ferroelectricity.[132–135] Much of this success comes from the high quality graphene/hBN interface resulting from the inert and atomically smooth hBN surface relatively free of dangling bonds and charge traps. This has naturally motivated studies exploring its suitability for 2D TMDs.[136–138] However, its small dielectric constant ($\kappa_{hBN}$ 5) means that even in perfectly crystalline, defect-free hBN, gate current leakage levels are predicted to be too large for its use as a gate dielectric in ultra- scaled devices with <1 nm EOTs.[139]

Other crystalline dielectrics used with 2D TMDs have been reported including mica,[140, 141] $CaF_2$,[59, 60, 62] $MoO_3$,[142, 143] VOCl,[144] $MnAl_2S_4$,[145], molecular $Sb_2O_3$,[146] and $SrTiO_3$ perovskites.[63, 64] However, more work is required to fully evaluate these materials beyond proof-of-concept devices. For example, while the initial demonstration of a top-gated device with mica exhibited low $SS = 72$ mV/dec and low $D_{IT} = 8.8 \times 10^{11} cm^{-2} eV^{-1}$,[140] subsequent measurements showed that long-term stability of mica encapsulated devices were poor due to the strong hydrophilism of mica trapping adsorbed moisture at the interface.[141]

Some promising materials to emerge in recent years are $CaF_2$[59, 60, 62] and $SrTiO_3$.[63, 64] As a dielectric, $CaF_2$ has a relatively high dielectric constant of 8.4 and a wide band gap of 12.1 eV (Figure 3f).[60] More importantly, its highly crystalline nature and a surface terminated with inert fluorine atoms allow well-defined interfaces with 2D TMDs. $MoS_2$ back gate FETs created with epitaxial $CaF_2$ dielectrics 2 nm thick exhibited competitive device performance, with $SS$ down to 90 mV/dec, small hysteresis, low gate leakage currents and current on/off ratios of $10^7$. $CaF_2$ belongs to a class of fluoride dielectrics that also include $SrF_2$ and, $LaF_3$, $MgF_2$, $BaF_2$ and others, which might also be interesting to explore. Likewise, the perovskite $SrTiO_3$ is a highly crystalline ultrahigh-$\kappa$ ($\kappa_{SrTiO_3}$) material that has been demonstrated as back gate dielectrics in $MoS_2$ FETs, achieving $SS$ down to 66 mV/dec and current

on/off ratios of $10^8$ (Figure 3g-j).[63]

However, a key limitation of many crystalline materials is the stringent growth requirements. The highest quality hBN are grown at demanding temperatures and pressures reaching 1500 °C and 4.5 GPa.[147] Crystalline hBN grown by CVD do not reach similar quality levels, lacks thickness tunability, and requires specially prepared substrates.[46] Indeed, epitaxial growth of single crystals is necessarily determined by the template substrate; this is also true for $CaF_2$ and $SrTiO_3$. Scalable transfer techniques will need to be developed, otherwise devices are limited to back-gated geometries. Polymer-assisted transfer of $SrTiO_3$ has helped realised proof-of-concept top gated devices but remains unscalable.[63, 64]

### 6. Native oxides

Native oxides refer to films oxidized on a semiconducting surface. In silicon, the quality and stability of its native oxide has underpinned its success over other superior channel materials and led to its dominance in modern electronics. Extending this approach to 2D semiconductors could lead to conformal synthesis of ultrathin, high-quality oxides with atomically flat and lattice-matched interfaces with low trap densities. Studies have been reported for $WSe_2$,[35, 148] $MoS_2$,[36] $TaS_2$,[37] $HfS_2$,[38, 39] $HfSe_2$,[40] and $ZrSe_2$.[40] Most of these early attempts produced amorphous, non-stoichiometric oxides and defective interfaces. However, a promising UV-assisted process enabled wafer-scale, polymer-free and area-selective oxidation of a non TMD 2D semiconductor.[41, 42] The resulting highly crystalline oxide $\beta$-$Bi_2SeO_5$ is high-$\kappa$ ($\kappa_{Bi_2SeO_5}$ =22) and compatible with a selective etching process. Their device achieved an EOT of 0.41 nm, low gate leakage current levels below the low power limit of 0.02 $Acm^{-2}$ and SS of 65 mV/dec with small hysteresis. While promising, the applicability of this approach to a broad range of 2D semiconductors including TMDs is unknown and should be the next phase of study. This approach will be more challenging as the thickness of the 2D semiconductor approaches the monolayer limit; realizing the controllable production of a high-quality monolayer FET with native oxidation will be significant.

### 7. Outlook

No single dielectric meets all the requirements for 2D TMD nanoelectronics. Many well-developed dielectric integration techniques from industrial CMOS technologies are not easily transferable to 2D TMDs. We must categorize dielectrics beyond material types, as materials can have significant differences in quality, compatibility and scalability depending on the integration or growth technique. For example, high quality hBN is only available through mechanical exfoliation, which is not scalable, while scalable CVD hBN is more defective. The high variance in reported material quality, of both the dielectric and 2D semiconductor, makes it difficult to select strong candidates. Numerical simulations have attempted to address this problem. However, such simulations are typically performed for defect-free cases, and usually without consideration of the integration feasibility with 2D materials. For example, comparisons of experiments and best-case simulations can result in gate leakage current estimates that span three orders of magnitude.[8] More simulations incorporating realistic defects will be needed to address these gaps.

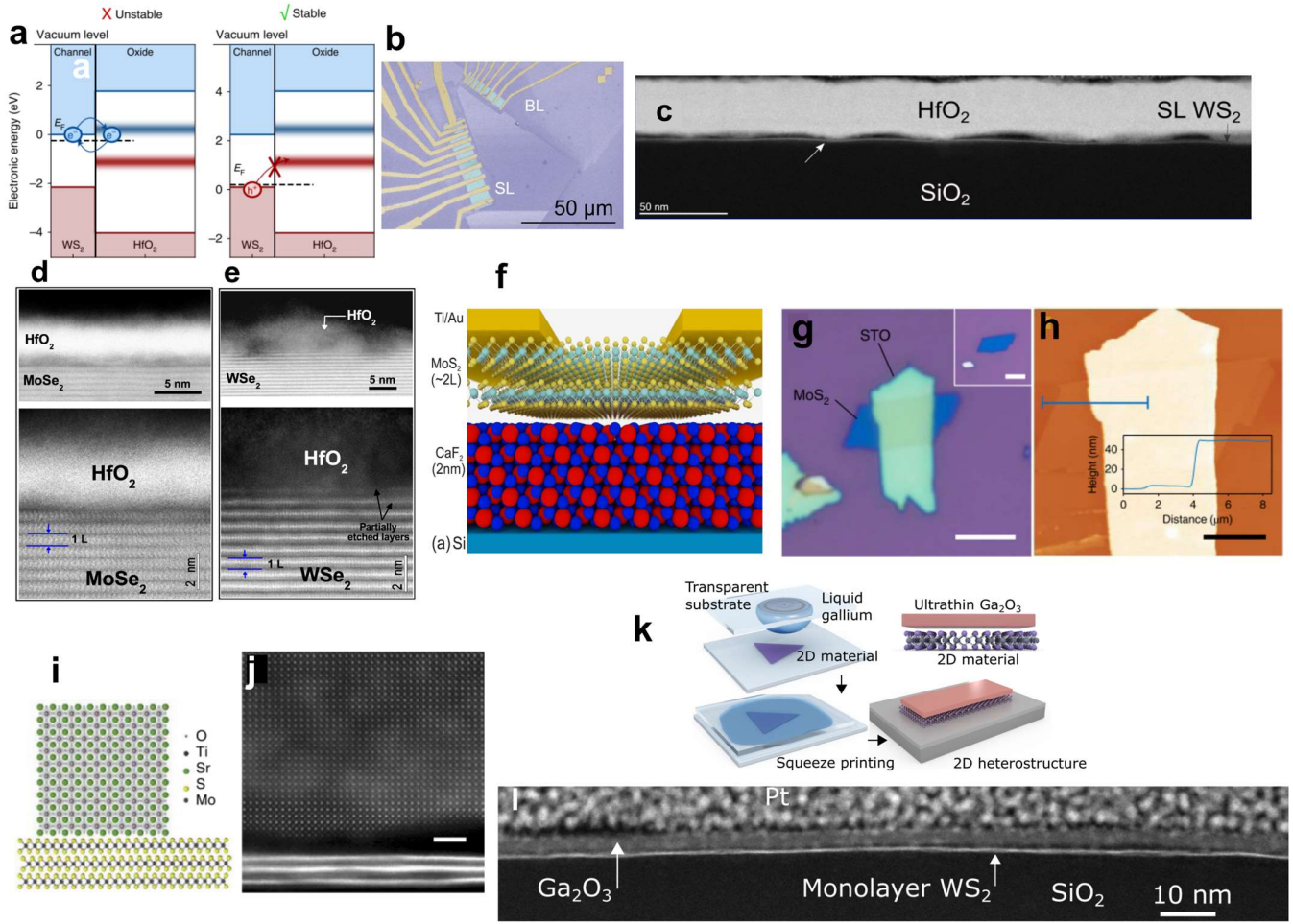

FIG. 3. (a) Band diagram schematic showing how device stability can be improved by Fermi-level tuning. When the Fermi level is aligned close to the conduction band edge (valence band edge), the device is unstable (stable). (b) Optical and (c) cross-sectional scanning transmission electron micrograph of 20 nm atomic layer deposition grown $HfO_2$ on 2D $WS_2$ revealing a rough interface that degrades carrier mobility. Cross-sectional images of $HfO_2$ growth on UV-$O_3$ treated (d) $MoSe_2$ and (e) $WSe_2$. 1L = one layer. Such surface treatment can damage the top-layers of the 2D materials as shown in the partially etched layers. (f) Schematic of a $CaF_2$ back gate bilayer $MoS_2$ device. (g) Optical image of a $SrTiO_3$ thin film transferred over an exfoliated $MoS_2$ flake. (h) Atomic force micrograph of the flakes, with the height profile shown in the inset. (i) Schematic of the crystal structures of $SrTiO_3$ and $MoS_2$. (j) Cross-sectional scanning transmission electron micrograph of the $SrTiO_3/MoS_2$ interface. (f) Schematic of a liquid metal printing process for ultrathin $Ga_2O_3$ on 2D TMDs, leading to atomically smooth interfaces as shown in the (g) cross-sectional transmission electron micrograph. (a) Reprinted with permission under a Creative Commons Attribution 4.0 International License from ref [106]. Copyright 2022 Springer Nature. (b, c) Reprinted with permission from ref [90]. Copyright 2021 John Wiley and Sons. (d, e) Reprinted with permission under a Creative Commons Attribution 3.0 Licence from ref[52]. Copyright 2015 IOP Publishing. (f) Reprinted with permission under the Creative Commons Attribution 3.0 Licence from ref [59]. Copyright 2015 IOP Publishing. (g-j) Reprinted with permission under the Creative Commons Attribution 4.0 International License from ref [63]. Copyright 2022 Springer Nature. (k, l) Reprinted with permission from ref [129]. Copyright 2023 American Chemical Society.

Other interesting device geometries have also garnered interest for surpassing the SS room temperature thermionic limit (sometimes referred to as the 'Boltzman tyranny'), such as tunnel FETs,[149] strain effect transistor,[150] and negative capacitance FETs (NC-FETs)[151–156] While some fundamental concepts on the nature of negative capacitance are still under debate,[157, 158] NC-FETs have nevertheless attracted substantial interest as device geometries can be similar to conventional MOSFETs and achieved by simply integrating an additional ferroelectric dielectric into the gate stack.[155, 159] Interest was further catalysed by the unexpected discovery of ferroelectricity in $HfO_2$ and $ZrO_2$ based thin films, industrial-compatible materials already used for CMOS and dynamic random access memory (DRAM) applications.[160, 161] Various device functionalities have been demonstrated for 2D materials integrated with ferroelectric gate dielectrics[162] that can potentially overcome the 'Boltzman tyranny' including polymer poly(vinylidene fluoride-trifluoroethylene) [P(VDF-TrFE)] with $Al_2O_3$,[152] pulsed laser deposition hafnium zirconium oxide (HZO) with $HfO_2$,[163], ALD HZO with $HfO_2$,[153] and ALD HZO with $Al_2O_3$,[154, 164]. Another promising near-term application of 2D transistors is ferroelectric-FET (Fe-FET) low-power, non-volatile memory technology that potentially of- fers fast switching speeds, long retention times, and high reliability compared to current floating-gate flashdevices. 2D memory Fe-FETs have been demon- strated with several types of ferroelectric gate di- electrics including P(VDF-TrFE),[165] epitaxial single crystal $PbZr_{0.2}Ti_{0.8}O_3$,[166] $CuInP_2S_6$,[167] and ALD $Hf_{0.5}Zr_{0.5}O_2$.[168] BEOL compatible Fe-FET memories with processing temperatures below 350 °C have also been demonstrated using AlScN.[169]

For such multilayers, manufacturing compatibility are key as they need to survive multiple potentially harsh processes or high temperatures. Materials should be chemically, electrostatically, and thermally stable. Furthermore, reliability considerations may be more complicated in gate stacks with V-shaped field profiles and additional dielectric/ferroelectric interfaces, which can lead to BTI, leakage, and breakdown issues.[158, 170, 171] Integrating sufficiently thick ferroelectric layers into a gate stack while ensuring good capacitance matching, high speed and reliability is necessary for CMOS thickness scaling. Ferroelectric materials typically have suppressed ferroelectricity at the few-nm regime, and the ferroelectric memory window is proportional to the ferroelectric thickness and coercive field. Here again, $HfO_2$ based ferroelectrics may be promising given that ALD grown ultrathin $HfO_2$ films down to 1 nm thick can exhibit ferroelectricity.[172]

Future experiments must explore beyond the dielectric itself and provide a holistic assessment of dielectric/2D semiconductor heterostructures and devices. Beyond the exploration of different dielectric integration strategies, experiments to reveal fundamental, atomistic level insights into dielectric and 2D semiconductor interactions will be valuable. For example, experiments revealed the origin of Fermi-level pinning in 2D semiconductor contacts as due to material damage from metal evaporation.

This motivated the search for low-melting point metals that can be evaporated gently, e.g., indium,[33, 173] bismuth,[174] and selenium.[175] Such knowledge can be employed for dielectrics as well, e.g., using materials that can be deposited with minimal damage to 2D semiconductors as seed layers for ALD.

Non-conventional ion gel electrolytes have also been explored for ion-gated $MoS_2$ vertical FETs [66] and $ReS_2$ FETs [67]. Their high specific capacitance enables energy-efficient, low-power applications. Devices can exhibit ambipolar behavior with high carrier densities and good transport properties such as high current densities (>3000 $Acm^2$), carrier mobilities (58 $cm^2V^{-1}s^{-1}$), on/off ratios ($10^6$), and low subthreshold swings (200 mV/dec). Another intriguing prospect is liquid metal synthesized ultrathin oxides,[70, 71] of which several have been demonstrated including high-$\kappa$ dielectrics such as $Ga_2O_3$,[70, 176] $HfO_2$,[70] $Al_2O_3$,[70] and $Gd_2O_3$[70]. Liquid metal synthesized oxide was demonstrated as a gate dielectric in 2D TMDs.[129] Here, an ultrathin $Ga_2O_3$ layer was directly printed on 2D $WS_2$ from liquid gallium in a low-temperature process and led to atomically smooth interfaces (Figure 3k, l). Subsequent ALD growth of $HfO_2$ led to a gate dielectric stack that achieved 1 nm EOT with excellent subthreshold swings and gate leakage currents. The realisation of $HfO_2$ based nanoelectronics will be strongly dependent on the community's exploration of more dielectrics and dielectric integration approaches.

## IV. OPTOELECTRONICS

2D materials have emerged as strong candidates for optoelectronics. While graphene is commonly used as a broadband absorber and in high-speed applications, its lack of a bandgap results in high dark currents and poor electrostatic control. This limitation can be circumvented in 2D TMD semiconductors.[177, 178] 2D TMDs are useful for optoelectronic applications as they can address a wavelength range ($\sim$1-2.5 eV) including the visible, have higher absorption (from $\approx$ 10% in monolayer to $\approx$ 90% in multilayer [179–182]), and have excellent gate tunability. Optical properties of 2D TMDs in visible range are dominated by sharp excitonic resonances which are extremely sensitive to the dielectric environment [183–188], especially in the monolayer form. Therefore, the dielectric can influence the performance of optoelectronic devices. Here, we begin by discussing the dielectric influence on 2D TMD excitons, before focusing on dielectric considerations for 2D TMD-based optoelectronic and photonic applications.

### A. Excitons in 2D TMDs

Excitons are bound quasi-particles formed via Coulomb interaction of electrons and holes. Unlike in conventional 3D quantum wells where excitons exist only at low temperatures, 2D excitons in monolayer TMDs are formed with extremely large binding energies (BE > 0.5 eV) due to reduced screening in 2D materials,[179, 183, 189, 190] promising for room-temperature optoelectronic applications. Figure 4a schematically shows that 2D TMDs on a Si/SiO$_2$ surface can undergo fluctuations in the dielectric environment from surface/interface irregularities of the dielectric. These irregularities can strongly influence the exciton binding energy and the lifetime $\tau_{life}$. An important criteria is therefore that TMDs are placed on (between) dielectrics with defect-free, atomically smooth interfaces. So far, hBN is the most successful candidate for these considerations, but remains limited to unscalable exfoliated flakes. Besides hBN, other dielectrics such as MoO$_3$ [191], Ga$_2$O$_3$ [176], Gd$_2$O$_3$ [192], and talc (Mg$_3$Si$_4$O$_{10}$(OH)$_2$)[193] have shown initial potential as dielectrics for exciton device applications, but further experimental investigations are needed. We highlight that there are different requirements for exciton recombination based on the application. A high-$\kappa$ dielectric (e.g., Al$_2$O$_3$[194], HfO$_2$[194, 195] etc.) prolongs excitonic lifetimes which is desired for photodetectors, while low-$\kappa$ dielectrics are suitable for photoemitters where faster recombination rates are preferred.

Beyond sandwiching the 2D TMD between two dielectric layers, integrating more layers to realize multilayered excitonic quantum well structures can unlock opportunities for engineering optical dispersions.[196] This can lead to exotic applications such as atomically thin electro-optical modulators and non-linear photonic devices. As compared to conventional devices with stringent requirements for the lattice-matched growth of each subsequent layer, 2D TMDs can instead be deterministically stacked and assembled into vdW layered heterostructures without such growth constraints. For example, uniform wafer scale growth permitting light trapping at excitonic resonances by engineering the surrounding dielectric environment was demonstrated.[196] In this work, CVD grown hBN and ALD grown Al$_2$O$_3$ dielectrics were used where the specific requirements were low-loss, uniform wafer-scale growth, precise thickness control, and high-$\kappa$.

Integrating TMDs into functional photonic devices such as coupling with an optical cavity engineered from assembly of dielectric layers could enable strong light-matter interactions to realize applications such as low-power optical switches, and enhanced photovoltaics and photodetectors. There are excellent reviews on strong light-matter interaction of 2D materials for various applications.[197, 198]. Here, we focus on the dielectric requirements which include: (i) uniform and precise thickness and roughness control as excitonic resonance strongly depends on the dielectric environment. (ii) High refractive indices for efficient light trapping [199, 200]. (iii) Low-loss to enable high cavity quality factors for enhanced emission. Another area of growing interests is the strain-induced, localized exciton in TMDs which has potential applications in quantum information processing.[201] Here the dielectric acts as a spacer or insulating layer for electrical gate control. Additionally, 2D dielectrics such as hBN can also act as room temperature single photon emitters.[202, 203]. Dielectric integration with single photon emitters is an ongoing effort due to its potential as quantum sensors and quantum information processing platforms.

Excitons in TMDs continue to be an active field of research with many applications predicted including hybrid metasurfaces and atomically thin flat optics for various domains ranging from quantum science to augmented reality, free-space optical communication [113, 204]. Most efforts so far have focused on the active TMD material and device integration. It is likely that more efforts on understanding compatible dielectrics will be key to realizing higher device performance.

## B. Photodetectors

Optoelectronic devices can be broadly categorized as photodetectors and photoemitters. While photodetectors convert optical signals into electrical signals, the electrical pumping in photoemitters produces an optical signal. In photodetection, carriers should be collected before they recombine i.e., a low recombination rate is preferred. In contrast, photoemitters require efficient exciton radiative recombination for light emission. Since the dielectric plays a key role in determining the exciton lifetimes, its choice can determine the suitability of the device for detection/emission purposes.

In general, a top dielectric for photodetectors is mainly for environmental passivation for increased device stability.[194, 205] In such cases, the top dielectric should be transparent across the wavelength of interest and be easily integrable with 2D TMDs without damage. Both bulk and interface defects need to be low as they can introduce mid gap states that absorb photons and lower the photoresponsivity (ratio of photocurrent to incident light power), mobility and internal quantum efficiency (photocurrent generated by absorbed photons). For gated $p$-$n$ junctions, the main requirement is a high capacitive coupling, which places stronger demands on high-$\kappa$ and thin dielectrics.

The two most common photodetector types are photodiodes and photoconductors. In a photodiode, a built-in potential allows the photogenerated electrons and holes to move and be collected at opposite contact electrodes. In photoconductors, the photoexcited electrons and holes are instead separated by applied external field. Hence carrier mobility determines the transit time and speed of the device. A high-$\kappa$ dielectric can screen Coulombic impurities to preserve high carrier mobility [206]. The high capacitive coupling also enables sharp $p$-$n$ junctions generated with external bias in a split gate structure [207] (Figure 4b). High carrier mobility and efficient gating enable high-speed, low-power, high-sensitivity photodetectors [208]. They therefore share many similar dielec-tric requirements discussed in the Nanoelectronics section. Smooth surface roughness is also critical, not only to maintain high carrier mobility, but also because of dielectric fluctuations from surface roughness.

Photodetector performance can also be increased via the photogating effect, where large photocurrents are generated by surrounding a 2D TMD active layer in an environment containing a high density of trap sites, i.e., the dielectric or interface (Figure 4c).[209] Photogating extends carrier lifetimes and increases photoconductive gain. The signal-to-noise ratio (SNR) can be effectively tuned by the electrostatic potential across the dielectric, which improves the sensitivity and bandwidth product of gain [177, 210]. However, such devices suffer from a slow response time; there is a trade-off between gain and speed. Dielectrics can also be useful to passivate the active layer from the environment or to improve speed, responsiveness and suppress hysteresis. The choice of dielectric, ranging from hBN, $SiO_2$, $HfO_2$, SiN etc. [194, 195, 206, 211] can therefore be engineered to provide specific functionalities depending on photodetector specifications.

### C. Photoemitters

Similar to photoluminescence, layered materials are excellent candidates for electroluminescence where photon emission is generated by electrical excitation. Strong light-matter interaction and a direct band gap in monolayer TMDs are promising for light-emitting diodes (LED) or single photon emitters [213, 217, 218]. Their strong optical absorption at excitonic resonance leads to $> 10^4$ higher quantum efficiency compared to multilayer TMDs[179]. The excitonic absorbance can be efficiently tuned to a narrow bandwidth via dielectric encapsulation.[219] However, the planar structure of the $p$-$n$ junction suffers from non-radiative recombination losses and poor in-plane mobility of the charge carriers. [220, 221] Therefore, vertical heterostructures have attracted attention as LED devices[213, 217]. In vertical heterostructures, the $p$ and $n$ type materials are stacked to increase the overlap emission area with the top and bottom electrodes. However,TMD/TMD heterostrucrures exhibit fast charge transfer (on the order of ~150 fs), thus hampering the LED performance. This can be improved by introducing a thin dielectric layer between top and bottom TMD, so that the charge carriers can tunnel from the top to bottom layer, increasing electroluminescence efficiency[222]. The main requirement is an ultra-thin dielectric that preserves the interfacial and material quality. Currently, hBN is widely used due ease of integration with TMDs and the possibility to exfoliate monolayer single-crystalline films of 0.33 nm in thickness. Thus, hBN improves LEDs in two ways: (i) as a dielectric in a split-gate structure to form a $p$-$n$ junction (Figure 4d), and (ii) as a thin tunnel barrier that allows current flow in a vertical heterostructure [223]. Depending on the required function, thin dielectrics with high-$\kappa$ can tune carrier generation by reducing exciton BE or carrier lifetime. A thin dielectric can also help to realize multilayered quantum well structures by further increasing the quantum efficiency [213] (Figure 4e).

### D. Outlook

Because of their excellent gate tuneability, TMDs can also be used as electro-optical modulators.[224, 225] For example, similar to a MOSFET configuration, excitonic emission can be tuned by varying the doping concentration in the material. Thus, by changing the gate bias to increase the ratio of charged excitons to neutral excitons, the oscillator strength can be increased. The dielectric here capacitively couples with the active material to modulate the emission spectra (Figure 4f). Optical modulators have been demonstrated with several TMDs where the reflectance at excitonic resonance can be efficiently tuned [215, 216, 226, 227]. When a vertical electric field is applied, exciton emission can show quantum confined Stark effect [214, 228]. However, this effect is small for intra-layer excitons because of their small out-of-plane polarizability[228] but can be significantly enhanced when an electric field is applied for hetero-bilayer system with inter-layer excitons[229, 230]. In these cases, a high-$\kappa$ dielectric is useful. While the carrier density ($n$) in the 2D active layer is proportional to the capacitive coupling ($n \propto C_{tot} V_g$ where $C_{tot}$ is the effective capacitance, and $V_g$ is the applied gate voltage), the applied electric field strength can be approximated as $F = \frac{V_g}{t_{2D} + t_d \frac{\epsilon_{\perp,2D}}{\epsilon_{\perp,d}}}$. Here, $F$ is the vertical electric field, $t_{2D(d)}$ is the thickness of 2D material (dielectric) and $\epsilon_{\perp,2D(d)}$ is the static out-of-plane dielectric constant of 2D material (dielectric).

Another feasible way of realizing high gain photodetectors is by integrating 2D TMDs with quantum dots by separating them with a buffer dielectric layer e.g., $TiO_2$ [231]. Additionally, pairing with the cavity mode through a photonic crystal cavity can enhance spin-polarized emission and subsequent confinement. A distributed Bragg Reflector increases the quantum efficiency drastically compared to placing it on conventional $Si/SiO_2$ substrates [232], representing a promising route towards onchip quantum information processing.

### E. Flexible Electronics

Flexible electronics are assembled on bendable, twistable, and stretchable substrates and have enabled technologies such as rollable displays, conformable sensors, printed radio frequency identification tags, and biocompatible electronics.[233–237] Owing to their high electronic performance and mechanical flexibility, 2D TMDs have potential applications as active layers in flex-ible electronics because of their small thicknesses.[1, 238–240] When bending a device stack, all layers are subjected to strain that is compressive at one side and tensile at the opposite. The maximum strain ($\epsilon$) can be expressed by $\epsilon = t/2r$, where $t$ is the stack thickness and $r$ the bending radius.[241] For the same strain, 2D materials with a smaller $t$ can therefore have a smaller $r$, i.e., they can be bent more with a larger curvature.

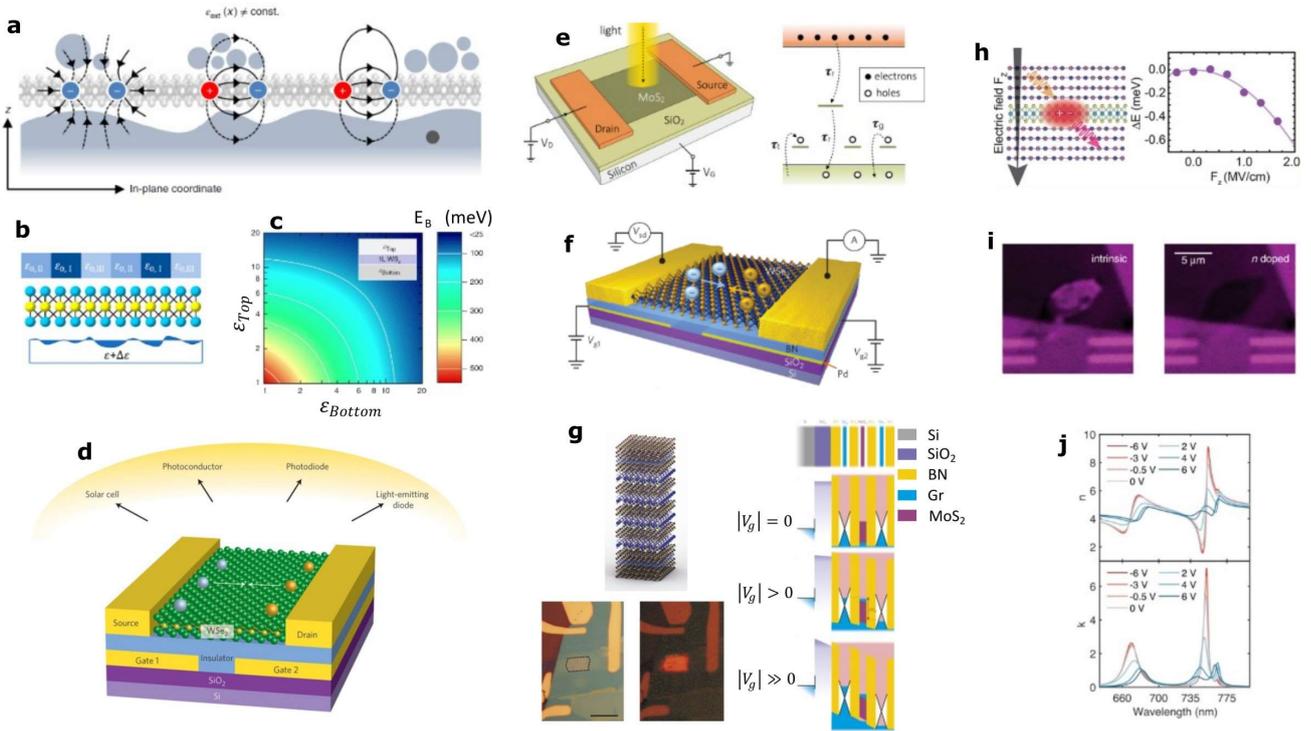

FIG. 4. Dielectrics for 2D TMD based optoelectronics. (a) Modulation of dielectric screening due to disorder induced by surface irregularities (top panel). (b) Schematic representation of a monolayer in between ordered and disordered dielectric environment. (c) Calculated BE modulation of exciton in monolayer $WS_2$ sandwiched between top and bottom [$\epsilon_{top(bottom)}$] dielectrics. (d) Schematic drawing of a lateral $WSe_2$ $p$-$n$ diode under split-gate geometry. The metallic back gates are separated from the $WSe_2$ monolayer by an insulating dielectric (SiN, $HfO_2$ or BN). Blue and orange spheres represent electrons and holes, respectively. (e) Schematic illustration of photogating effect where carriers are being trapped in the mid-gap state upon illumination where $MoS_2$ is placed on a $SiO_2$ layer. (f) Schematic of multilayer $WSe_2$ $p$-$n$ junctions with palladium back gates ($V_g$). During electroluminescence in $WSe_2$, electrons (blue) and holes (yellow) move towards each other (arrows) and recombine. (g) Schematic of multiple quantum well heterostructures (individual layers are colour coded) and corresponding optical micrographs of the operational electroluminscent device at room temperature. Band diagrams for the case of zero applied bias ($|V_g| = 0$), intermediate applied bias ($|V_g| > 0$) and high bias ($|V_g| \gg 0$) for the heterostructure presented in the top panel. Optical modulators demonstrating quantum confined Stark effect (h), optical mirror (i) and modulation of $n$ and $k$ values under $V_g$ (j). (a) Reprinted with permission from ref[183]. Copyright 2019 Nature Publishing Group. (b) Reprinted with permission from ref[185]. Copyright 2021 American Chemical Society. (c) Reprinted with permission under a Creative Commons Attribution 4.0 International License from ref[184]. Copyright 2017 Nature Publishing Group. (d) Reprinted with permission from ref.[207]. Copyright 2014 Nature Publishing Group. (e) Reprinted with permission from ref.[209]. Copyright 2014 American Chemical Society. (f) Reprinted with permission from ref.[212]. Copyright 2014 Nature Publishing Group. (g) Reprinted with permission from ref.[213]. Copyright 2015 Nature Publishing Group. (h) Reprinted with permission from ref[214]. Copyright 2015 American Chemical Society. (i) Reprinted with permission from ref [215]. Copyright 2018 American Physical Society. (j) Reprinted with permission from ref [216]. Copyright 2021 American Chemical Society.

Besides the active layer, other key components, such as electrode, gate dielectric, and substrate, must also be compatible for flexible electronics. Flexible electronics share many similar dielectrics requirements as those dis-cussed in the Nanoelectronics section. Additionally, they should be grown at low temperatures as most flexible materials will be deformed at elevated temperatures. The device performance must also be maintained when bending with large strain.

The flexibility of a solid material under uniaxial stress ($\sigma$) can be expressed by $E_Y = \sigma(\epsilon)/\epsilon = FL_0/A\Delta L$, where $E_Y$ is the Young's modulus, $F$ the force exerted on the material, $A$ the cross-sectional area perpendicular to the applied force, $\Delta L$ length change of the material, and $L_0$ the original length of the object.[247] $E_Y$ is a fundamental material property; the higher the modulus, the more stress is needed to create the same amount of strain. Theoretically, a rigid body should have an infinite $E_Y$, while a soft material should have zero $E_Y$ (Figure 5). Polymers are frequently used as substrates and dielectrics, metals and alloys as electrodes, and oxides as dielectrics and electrodes. Although $E_Y$ of metals and oxides are higher than polymers, thin metal and oxide films can also be bendable. Here, we review three most attractive types of gate dielectrics for flexible electronics: polymer, amorphous oxide, and hybrid.

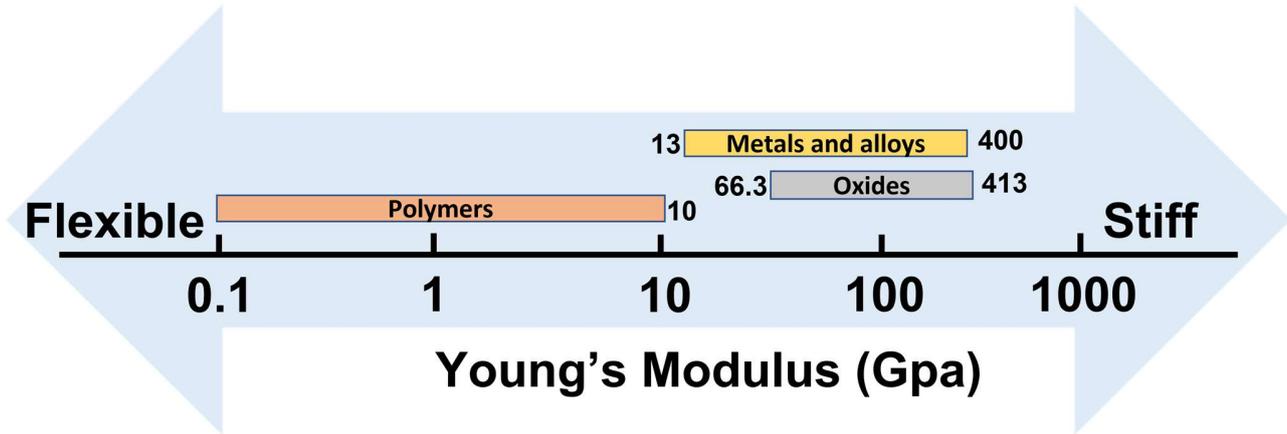

FIG. 5. Young's modulus spectrum of different materials describing their mechanical performance.[242–246]

### 1. Polymer Dielectrics

Polymers are widely used as dielectrics for flexible FETs due to their low processing temperature, high flexibility, and versatility. Polymer films can be formed at room temperature by methods such as solution processing, inkjet printing, spin coating, and ion-gel etc.[249–252] Although polymers have extremely small $E_Y$, they typically have low $\kappa$ and capacitance which are inefficient for power consumption. Therefore, achieving high-capacitance polymers is key. For conventional polymers, which generally have a lower $\kappa$ than $SiO_2$ ($\kappa$ = 3.9), various approaches have been proposed to increase their capacitance, such as polymer dielectric cross-linking,[253–255] high-$\kappa$ and low-$\kappa$ polymer blend dielectrics,[256–258] bilayer polymeric dielectrics,[259–261] and decreasing dielectric thickness.[262, 263] Many of these methods have increased capacitance of low-$\kappa$ polymers to over 10 nF/cm².

Fluoropolymers such as polyvinylidene fluoride (PVDF, $\kappa$ = 10.3),[264] poly(vinylidene fluoride-hexafluoro propylene) (P(VDF-HFP), $\kappa$ = 11),[265] and poly(vinylidene fluoride-trifluoroethylene) (P(VDF-TrFE), $\kappa$ = 15),[266] have relatively higher $\kappa$ due to the fluorine polarity and electronegativity in the polymer backbone. They are promising gate dielectrics for flexible FETs. For example, P(VDF-TrFE) films achieved a low leakage current density < $10^{-6}$ A cm$^{-2}$ at a bias of 20 V and a large breakdown strength of 2.5 MV cm$^{-1}$.[267] This insulating performance of P(VDF-TrFE) film is comparable with oxides which generally exhibit better insulating performance than conventional polymers.

Solid polymer electrolyte is another attractive option for high performance flexible 2D TMD FETs.[268, 269] Their key feature is a high specific capacitance from the formation of a sub-nm thin electric double layer at the electrolyte-semiconductor interface when the electrolyte is polarized. With a positive gate bias, anions in the electrolyte move towards the gate/electrolyte interface, while cations move towards the electrolyte/semiconductor interface. Subsequently, electrons accumulate in the semiconductor. For example, a $MoS_2$ flexible FET was realized by ion gel with an electrolyte dielectric drop-casted from an ethyl propionate solution of a triblock copolymer (PS-PMMA-PS) and [EMIM][TFSI].[65] (Figure 6a). The measured capacitance of the electrolyte films on the polyimide (PI) substrate is 4.67 μF cm$^{-2}$, which exceeds the capacitance requirement of 3 μF cm$^{-2}$ for nanoelectronics (Figure 6b). Moreover, the device performance under various bending conditions reveals negligible performance degradation during bending and good performance recovery after bending. Figure 6b shows the transfer curves after continuous bending ($r$ = 0.75 mm). The dependence of the drain current and normalized mobility as a function of bending radius $r$ is shown in Figure 6c. The performance variation under different bending conditions is less than 10 %. This work demonstrated the excellent electric and flexible performance of 2D TMD device based on polymer electrolyte dielectric.

Although most polymer dielectric materials meet the requirements of low temperature synthesis and mechanical flexibility of flexible FET, each material has its limit. Conventional polymer dielectrics still suffer from high leakage currents and relatively low capacitances. The ferroelectric property of high-$\kappa$ fluoropolymer dielectrics generates charge traps during polarization pro-ducing large transfer hysteresis.[256, 257, 270] As a result, high capacitances do not always lead to efficient charge carrier induction in the semiconductor. Solid polymer electrolytes also exhibit low thermal stability and solution-like property, leading to rough surfaces that make subsequent growth more challenging.[271]

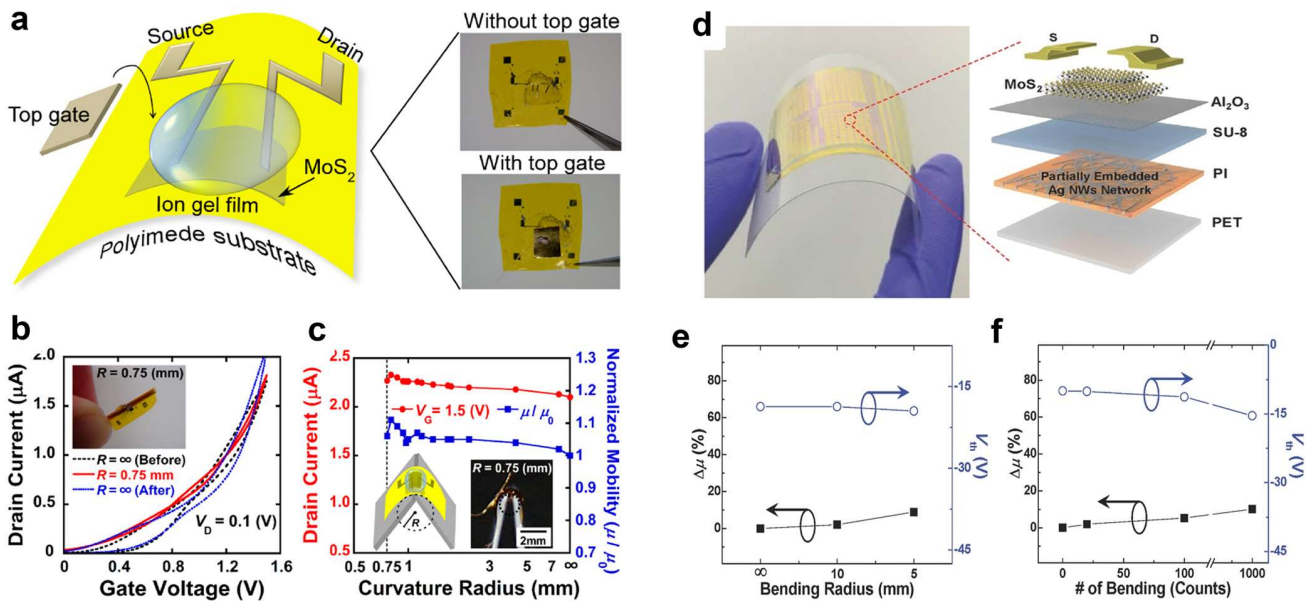

FIG. 6. A thin-film $MoS_2$ flexible field effect transistor (FET) constructed with a polymer dielectric on a polyimide substrate. (a) Schematic depiction of the $MoS_2$ FET. (b) Transfer characteristics of the $MoS_2$ FET. The red, black dotted, and blue dotted curves correspond to the transfer curve for a bending radius of 0.75 mm and to the transfer curves before and after the bending experiments, respectively. The inset shows the optical image of the 0.75 mm bending radius. (c) The dependence of the drain current (red) and the normalized carrier mobility (blue) on the bending radius at a gate voltage $V_G$= 1.5 V. The inset schematic illustrates the bending measurement. (d) Photograph of a bent flexible $MoS_2$ FET array. The inset shows the schematic layer-by layer structure of the flexible device. Change of mobility and $V_{th}$ of the flexible $MoS_2$ FETs as a function of (e) bending radius, and (f) bending cycles. (a-c) Reprinted with permission from ref [65]. Copyright 2012 American Chemical Society. (d-f) Reprinted with permission from ref [248]. Copyright 2016 John Wiley and Sons, Ltd.

### 2. Amorphous oxides

Amorphous oxides are attractive dielectrics for flexible 2D TMD devices with high areal capacitances and good mechanical flexibility.[272–274] As discussed previously, amorphous oxides are typically grown by processes like ALD. For flexible FET applications, the growing temperature cannot exceed the glass transition temperature of the flexible substrate. The glass transition temperature refers to the temperature at which a polymer changes between a hard and brittle 'glassy' state to a soft, viscous 'rubbery' state.[275] For example, PI and polyethylene terephthalate (PET) are widely used flexible substrates with glass transition temperatures of 250 and 140 °C respectively.[276, 277]

Amorphous $Al_2O_3$ ($\kappa$ = 8.5, $E_Y$, ~200 GPa) has a deposition temperature around 200 °C and is widely used as gate dielectrics for flexible 2D TMD FETs on PI substrates.[278–280] A flexible $WS_2$ FET based on ALD grown $Al_2O_3$ dielectric allowed small bending radius with modest stress on the device due to the thin and smooth PI substrate.[281] As a result, the $WS_2$ transistor survived 50,000 bending cycles between flat and $r$ =2 mm with an on-current reduction of only 30 %, indicating good flexibility of the device. The leakage current was also below $10^{-10}$ A throughout the bending test, demonstrating the good insulating properties of ALD grown $Al_2O_3$.

Amorphous oxide dielectrics have high $\kappa$ and good insulating properties allowing low operating voltages and high driving currents. Nevertheless, current methodologies to grow metal oxides are expensive, time-consuming, and sometimes incompatible with flexible substrate due to the high growing temperature. Recently, alternative low temperature fabrication of high-quality metal oxide dielectric films by solution processing method, such as DUV treatment, combustion synthesis, and liquid metal printing, show potential for dielectric applications in flexible 2D TMD FETs.[282–286]

### 3. Hybrid dielectrics

Hybrid dielectric materials have emerged as an alternative class of gate dielectrics. The concept is to combine and leverage the physical properties of organic polymer dielectrics and inorganic metal oxide dielectrics into a single component. Various hybrid dielectrics have been reported for flexible applications, including organometallic polymer dielectrics, organic-inorganic nanocomposite dielectrics, and bilayer organic-inorganic dielectrics.[287–292] For example, flexible $MoS_2$ FETs were fabricated with SU-8 and ALD grown $Al_2O_3$ bilayer gate dielectrics (Figure 6d).[248] The capacitance of the 630 nm-thick hybrid dielectric is 4.6 nF $cm^{-2}$, much higher than the capacitance of SU-8 alone.[293] Figure 6e, f shows the mobility and threshold voltage $V_{th}$ changes under various bending radius and bending cycles, respectively. The bent device showed negligible mobility and threshold voltage shift compared to the initial performance (less than 10 %). On the other hand, the mobility and $V_{th}$ is shifted about 10 % and 17 % after 1,000 bending cycles, respectively. Although the hybrid dielectrics are not as flexible as polymers, they have superior flexibility compared to oxide dielectrics.

While hybrid dielectrics can exhibit enhanced properties by leveraging the advantages of both organic polymer dielectrics and inorganic oxide dielectrics, they can also suffer from the associated limitations. These include increased fabrication cost, degraded flexibility compared to polymer dielectrics, and low $\kappa$ and weak breakdown field with respect to oxide dielectrics. More importantly, ultrathin multiple layer growth demands smooth film surfaces, which is a current bottleneck for solution process techniques.

### 4. Outlook

Future investigation of dielectrics for flexible 2D TMD FET application should focus on three approaches: developing alternative dielectrics, optimizing current dielectric growing process, and integrating dielectrics into 2D TMD transistors. The insulating properties of polymers can be easily tailored, and future developments should focus on dielectric materials/hybrid heterostructures with high $\kappa$, small $t$, and low leakage currents. Research on amorphous oxides should focus on developing recipes for high quality growth at low temperatures. Finally, many emerging dielectric materials with good insulating properties and flexibility have not yet been demonstrated in flexible 2D TMD FETs. Integrating these dielectrics with flexible 2D TMD FETs will be challenging but should allow for exciting functionalities and enhanced device performances.

Flexible products, such as rollable OLED TV, folding smartphones, and e-paper displays are already available in the commercial market.[294, 295] 2D TMD based products are yet to be commercialized, but expectations remain high.[296, 297] However, beyond dielectrics, other challenges such as contact engineering, material quality, scaling, and manufacturing process compatibility must be met before 2D TMDs can be used in commercial flexible applications such OLEDs and displays.[33, 298, 299]

## F. Biosensing

Field effect transistor-based biosensors (bio-FETs) detect biological molecules (target, target-analyte) which are usually biomarkers, i.e. their presence is indicative of a specific disease, contamination, or condition. Atomically thin 2D TMDs are promising candidates for next-generation bio-FETs due to their high surface-area-to-volume ratio and excellent electrostatics that facilitate highly sensitive bio-FETs.[300–303] Compared to graphene, the presence of a band gap allows for higher sensitivity. For sulphur based TMDs such as $MoS_2$ and $WS_2$, well-established surface functionalisation techniques can be used to achieve additional device sensing functionalities.

Bio-FET sensing can be performed in the following ways: measuring changes in the (i) threshold voltage $V_{th}$ or (ii) transconductance $g_m = \frac{\partial I_D}{\partial V_{GS}}$ from introducing a target analyte. (iii) Measuring drain-current $I_D$ changes in time as the target analyte concentration is varied (Figure 7a).[304, 305] Bio-FETs are made up of: (i) the receptor, a functional unit such as an antibody or aptamer covalently attached to the channel surface that selectively binds to the desired target-analyte. (ii) The transduction element and readout which convert the molecular binding event between target and receptor (biorecognition event) into a measurable electrical signal. (iii) A global back gate and source-drain electrodes.[306–308]

The role of the dielectric in 2D TMD-based bio-FET sensing is dependent on the device geometry. Bio-FETs can be constructed with: (i) only a back gate dielectric, or (ii) a dielectric-encapsulated-channel with both a back and top gate dielectric. In the latter, the dielectric screens charges within the channel and capacitively couples the sensing area to the device channel. The top dielectric also passivates the TMD channel to improve the air/solution stability of the device. It can also be an active grafting layer for the functionalisation of the device with the desired receptor unit. For example, APTES-gluteraldehyde modification is an oxide surface treatment for the subsequent covalent attachment of amine modified functional units.[311–313] Both device geometries allow measurements in a wet or dry state. Dry state sensing relies on device changes between sequential device measurements and electrolyte (typically aqueous bodily fluids such as sweat and saliva) introduction. Wet state sensing is a continuous measurement during electrolyte introduction. In wet state sensing, electrostatic screening from mobile ions in the electrolyte can reduce sensitivity. Though dry-sensing circumvents screening effects, extra processing steps, longer response times, and increased operational complexities are required.

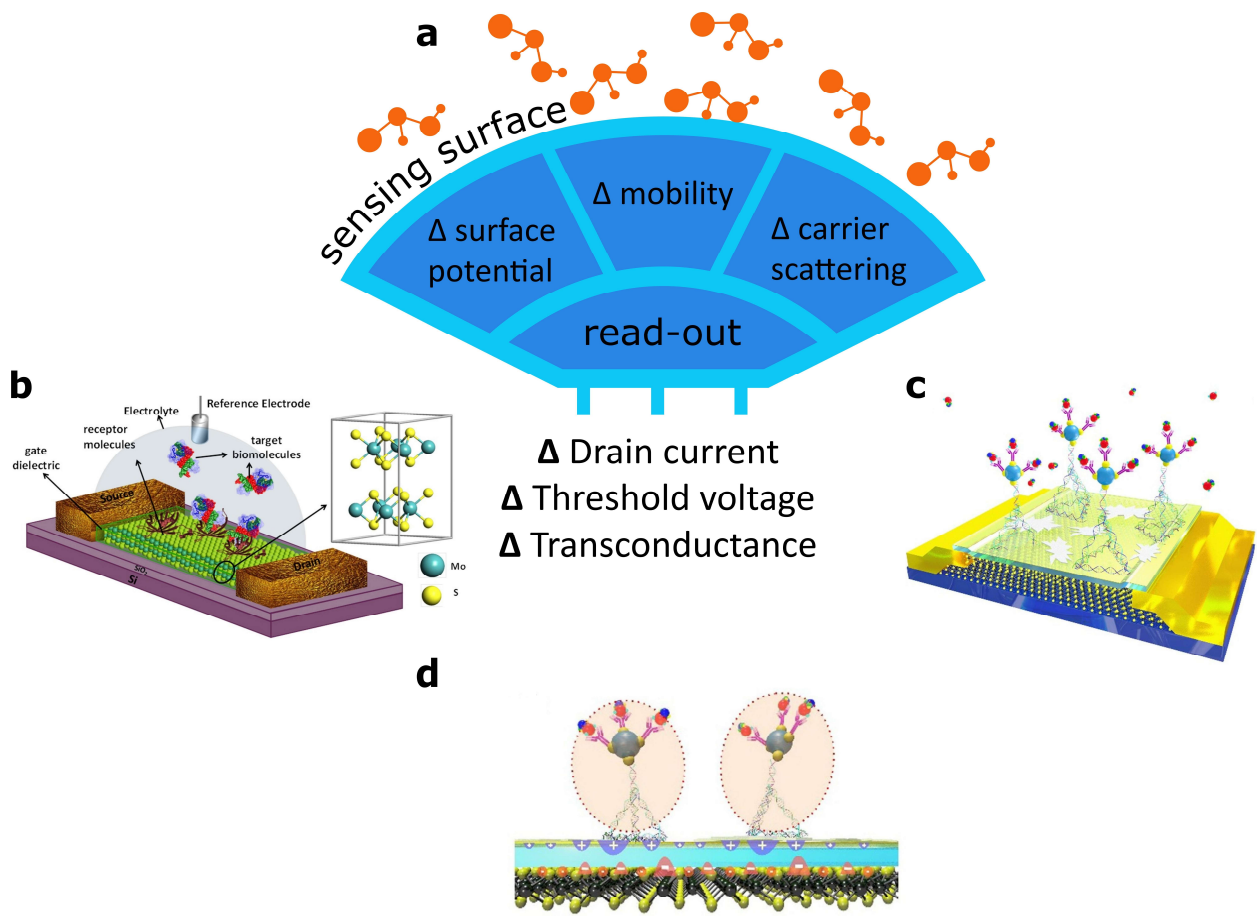

FIG. 7. (a) Schematic representation of FET based biosensing. Surface based biorecognition events result in changes to the transport properties of the underlying semiconducting channel, this can then be read-out as changes in the drain current $I_D$, threshold voltage $V_{th}$ and transconductance $g_m$ of the FET. (b) Device schematic of a MoS$_2$ bio-FET operated in wet-state. (c) Improved MoS$_2$ bio-FET employing thin and high-$\kappa$ dielectric (5 nm of Al$_2$O$_3$) and a DNA-tatrahedron based method of surface functionalisation. (b) Reprinted with permission from ref [309] Copyright 2014 American Chemical Society. (c-d) Reprinted with permission from ref [310]. Copyright 2021 Elsevier.

Dielectric selection in wet state bio-FETs can depend on the screening effect from mobile ions within the electrolyte, as described by the Debye length $\lambda_D = \sqrt{\frac{\epsilon k_B T}{q^2 I}}$ where $\epsilon$ is the relative permittivity of the solution, $k_B$ the Boltzmann's constant, $T$ the temperature, $q$ the electron charge, and $I$ the ionic strength of the electrolyte. In a wet state bio-FET, a sensed charge at the channel-electrolyte interface causes the migration of counter-ions within the solution toward the channel surface to create an electric double layer. This double layer in turn screens the channel surface and decreases device sensitivity. The screening distance is $\lambda_D$ which decays exponentially with increasing ionic strength of electrolyte. For reference, the average size of receptors is 10-15nm, while $\lambda_D$ in physiologically relevant samples (e.g., sweat and saliva) is 1 nm, which would lead to charge screening and ineffective biosensors. Consideration of $\lambda_D$ and receptor sizes thus determines the necessary dielectric scaling, i.e., their dielectric constants and thicknesses, to maximise capacitive coupling and stable passivation against air and solution-based degradation.[314, 315] In the following sections, the dielectric requirements of bio-FETs will be discussed with respect to scalability and quality.

### 1. Scalabilty

A reduced dimensionality can improve sensitivity .[316, 317]. For bio-FETs, decreasing free carrier screening and maximising capacitance are important dielectric requirements for effective sensing. The reduced free carrier screening intrinsic to 2D materials can be exploited for lower limits of detection, specifically in the subthreshold regime. A TMD based bio-FETs was demonstrated with few-layer $MoS_2$ and a 35 nm $HfO_2$ dielectric that served as an active grafting layer for receptor unit attachment (Figure 7b).[309] The sensitivity, defined as the normalised ratio of the difference in drain current before and after biorecognition event, was found to be 196 at 100 fM, the lowest concentration of analyte tested. Current state-of-the-art $MoS_2$ based bio-FETs can now achieve sensitivities of $\sim 50,000$.[310] As the key sensing mechanism of such bio-FETs involved capacitive coupling of the dielectric surface to the semiconducting channel, a thin and high-$\kappa$ dielectric would increase sensitivity.

More recent 2D TMD based bio-FETs have employed thinner high-$\kappa$ dielectrics and surface functionalisation strategies. Few-layer $MoS_2$ with 5 nm of ALD $Al_2O_3$ showed good air stability and resistance to liquid erosion.[310] Surface functionalisation of the $Al_2O_3$ layer with DNA-tetrahedron modified antibodies promoted biorecognition events and improved sensitivity (Figure 7c,d). The device achieved a limit of detection (LoD) of 1 fg/ml in both undiluted PSA and human serum which are both high ionic strength solutions.

### 2. Quality

Dielectric quality requirements for bio-FETs frequently mirror those of nanoscale FETs. Sensing performance is dependent on interface traps, dielectric defects, and surface roughness.[136, 318, 319]. Charge trapping/de-trapping at or near the interfaces will degrade the signal-to-noise ratio (SNR) and LoD. Low concentrations of target analytes that cause indistinguishable changes in $V_{th}$ or $I_D$ due to high background noise from these traps and defects will lead to ineffective sensing.

For bio-FETs with a top dielectric grafting layer, a clean and homogeneous dielectric leads to uniform and large-area receptor coverage. Biorecognition events can then act cohesively to gate the entire TMD channel, instead of acting as single point charges which would minimise scattering and degrade sensitivity (Figure 8a).[320] Removing the top dielectric layer can increase device sensitivity and simplify device fabrication and operation. However, it can also decrease device specificity and stability while requiring alternative receptor grafting strategies. For example, in $MoS_2$ bio-FETs without a top dielectric layer and no surface modification,[321] receptor attachment was instead promoted by the affinity of biological molecules towards the relatively more hydrophobic $MoS_2$ surface compared to the surrounding $SiO_2$ substrate (Figure 8b). Although the sensitivity was improved, the specificity suffered and the device gave unwanted responses to other molecules present on the channel surface.

### 3. Outlook

As healthcare management shifts towards early and effective intervention, rapid and accurate diagnostic tools are critical.[324] Cost-effective, wearable, low-power, user-friendly, and miniaturized point of care (PoC) devices (where testing occurs close to or near the patient rather than at a laboratory) can therefore be critical for early disease detection. Wet-sensing, with faster response times and reduced operational complexities may be more viable for such PoC applications [325–327] Bio-FETs with top dielectric layers are typically more suitable for PoC devices with faster sensing and increased air and solution stability.[310, 328–330] For increased power efficiencies, sensing mechanisms based on either (i) $V_{th}$ shifts or (ii) changes in $I_D$ with a constant $V_{DS}$ and $V_{GS} < V_{th}$ can increase sensitivity and allow for lower operating voltages.[331, 332]

Another key factor towards PoC applicability is the SNR. Amplifying the gate voltage applied to the channel by biorecognition events is an interesting strategy which can be achieved with NC-FETs.[151, 333] NC-FETs will require ferroelectric/dielectric heterostructures such as $HfZrO_2/Al_2O_3$[154] or $SrTiO_3/PbTiO_3$[334] and was recently demonstrated in $WSe_2/MoS_2$ ion sensitive field-effect transistors where the sensitivity to solution pH was improved from the 59 mVpH$^{-1}$ theoretical maximum to 362 mVpH$^{-1}$ [335]. Simulations showed that the addition of a negative capacitance Al-doped $HfO_2$ top-gate stack improved the voltage sensitivity by $\sim 100$ mVpH$^{-1}$ (Figure 8c). [322] NC bio-FETs are in the nascent stage of development and are an interesting avenue to improve device sensitives beyond current limitations.

Flexible and wearable electronics are other opportunities to integrate bio-FET technology into real-time sensing and monitoring of patient conditions. As outlined in the Flexible Electronics section, polymer composites can aid in the realisation of back-gated bio-FETs on flexible substrate/bottom-dielectric stacks. This was demonstrated with $MoS_2$ with PI as a flexible substrate and a hybrid back-gate dielectric composed of SU-8 and $Al_2O_3$ (Figure 8d).[323]

2D TMDs are promising for highly sensitive, fast, and reliable bio-sensors. The route towards commercial devices must consider their successful integration with various dielectrics to realise tailored functionalities.

### G. Quantum Information Processing

Quantum information processing (QIP) refers to the storing, processing, and transmitting of data impossible through classical means by leveraging quantum mechanical phenomena such as entanglement, superposition, and tunnelling.[336–338] The encoding of information bits in quantum two-level systems, known as qubits, is the basis of a quantum computer that can accelerate computational tasks intractable on classical computers. Physical qubits have been realised on many systems;[339] front runners include superconducting ciruits,[338, 340] colour centers,[341, 342] ion traps,[343–349] and semiconductor quantum dots.[350–353]

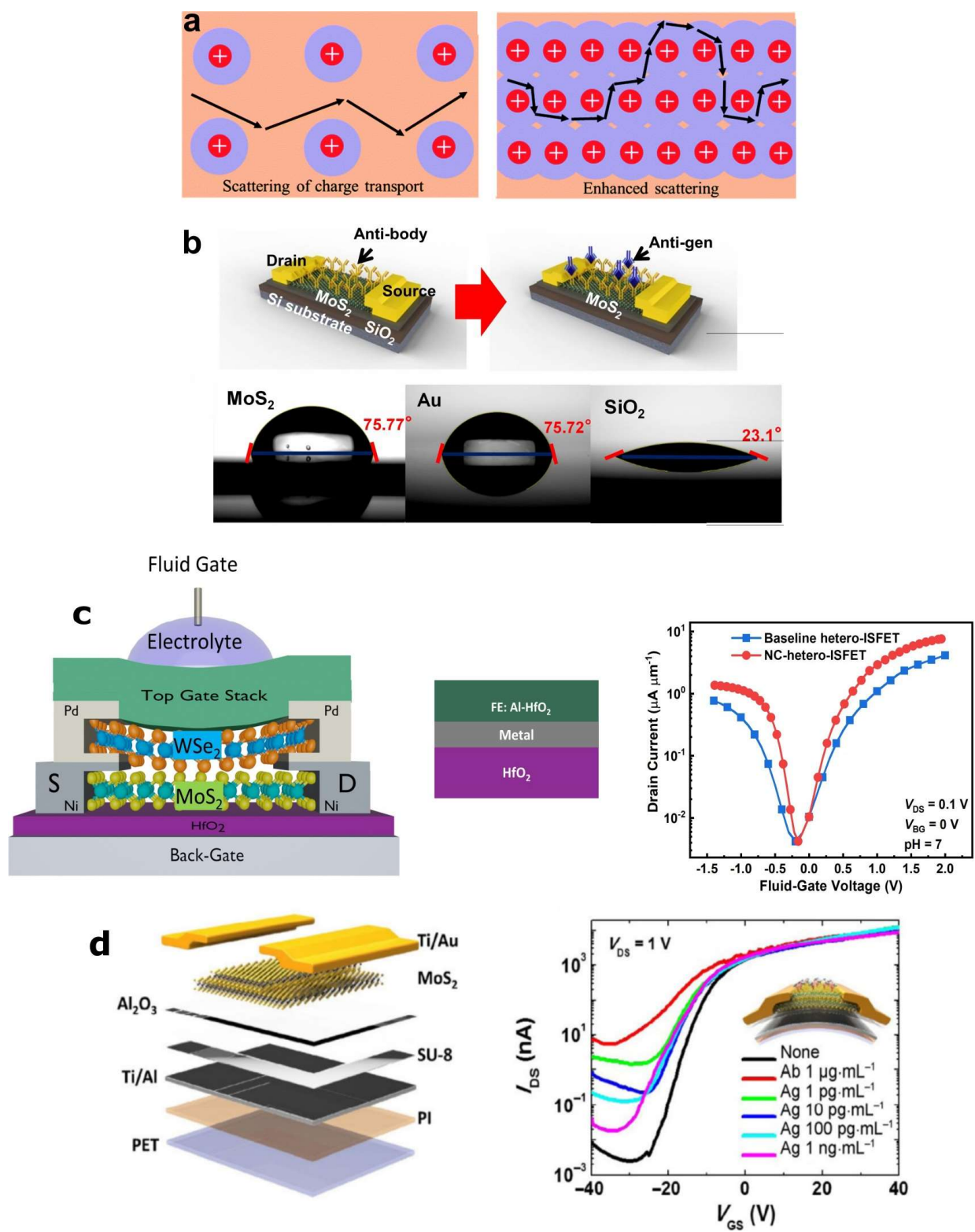

FIG. 8. (a) With increased receptor coverage, maximal carrier scattering and therefore sensitivity of the bio-FET can be achieved. (b) Dry-state bio-FET constructed without a top dielectric layer or surface modification to maximise sensitivity. Relative hydrophobicity was instead employed for surface modification with receptor units. (c) Device configuration of TMD heterostructure biosensor. Components of the dielectric stack modification employed to achieved negative capacitance. Resulting improvement in $I_D$ change for a given change in pH. (d) Schematic of MoS$_2$ based bio-FET constructed on a flexible substrate with an SU-8/Al$_2$O$_3$ bottom-gate dielectric stack for a balance between flexibility and high dielectric permittivity. Changes in $I_D$ demonstrated under flexing of the bio-FET as increasing concentrations of analyte are introduced to the channel.(a) Reprinted with permission from ref [320]. Copyright 2019 Royal Society of Chemistry. (b) Reprinted with permission under a Creative Commons Attribution 4.0 International License from ref [321]. Copyright 2014 Springer Nature. (c) Reprinted with permission from ref [322]. Copyright 2022 IOP publishing. (d) Reprinted with permission from ref [323]. Copyright 2017 Springer Nature.

In a gate-defined semiconductor quantum dot, charged carriers are spatially confined to a region $\sim$10-100 nm.[350, 351] This confinement is created by shaping electrostatic potential landscapes with gate electrodes and allow control over individual carrier degrees of freedom, such as the electron spin. Planar spin qubits are the most well-known implementation and have been demonstrated on various material platforms including III-V GaAs/AlGaAs heterostructures,[354, 355] Si,[356– 359] and Ge.[360–362]

Recently, 2D materials have emerged as qubit candidates; their atomically thin geometry presents natural carrier confinement along the out-of-plane dimension and improved electrostatics that facilitate carrier confinement.[363–369] While current state-of-the-art 2D material based quantum dot demonstrations are mostly on graphene,[370, 371] 2D TMD semiconductors offer many enticing upgrades.[372–379] 2D TMDs have direct band-gaps; unlike graphene which requires additional fabrication complexity to induce a bandgap, e.g., with atomically precise edges or perpendicular electric fields. Due to the $d$ orbital electrons in the metal atoms, TMDs have strong spin-orbit coupling useful for fast, all electrical qubit control via electric dipole spin resonance. The strong spin-orbit coupling also results in strong spin-valley locking with valley-dependent optical selection rules. These exotic properties can lead to hybrid spin-valley qubits offering fast optical/electrical control with long state coherence lifetimes, i.e., the time quantum information is preserved; fast control and longer lifetimes are useful for efficient QIP.[380–382]

Compared to other material platforms, the dielectric in 2D TMD quantum dots is arguably more influential due to their atomically thin geometry. The main role of dielectrics in semiconductor quantum dots is the electrical insulation of the active layer from the gate electrodes, and therefore shares many similar dielectric requirements with that of a FET, though sometimes for different purposes. For example, a small EOT in FETs enables fast and efficient device switching. However, a small EOT in quantum dot devices is critical to lithographically define small electrostatic potential landscapes ($\sim$10-100 nm) with steep slopes for effective carrier quantum confinement and manipulation.[383] This is especially important for TMDs with relatively larger effective masses $m_{eff}$ that demand small structure areas $A$ to achieve sufficient quantum level spacings $\Delta E$, given by $\Delta E = \frac{\pi \hbar^2}{m_{eff} A}$.

However, as TMD-based QIP is a nascent field, compatibility with scalable industrial processing is less important in the short term. The immediate aim is to demonstrate the feasibility of TMD qubits, e.g., few-electron regime, and effective state initialization, read-out and control fidelities. Quality considerations are more important. Qubits operate at cryogenic temperatures where carrier mobilities are dominated by impurity and interface roughness scattering. Low interface and bulk defect densities and atomically smooth interfaces are therefore crucial and contribute towards enhancing quantum state coherence lifetimes by minimizing charge noise, a major decoherence source.[351] Furthermore, atomically smooth interfaces reduce disorder from creation of random electron-hole puddles that distort potentials.[94, 384] Phonon scattering, on the other hand, is naturally suppressed due to cryogenic operation. High-$\kappa$ dielectrics may therefore work well, as they provide effective charge impurity screening while their disadvantage of introducing remote phonons is mitigated. Likewise, requirements on breakdown voltages, gate leakage currents, gate hysteresis, and operation lifetimes are relaxed as these typically improve at cryogenic temperatures.

We mention a dielectric requirement unique for QIP applications: potential for nuclear spin-free isotopic purification. Besides charge noise, a major source of decoherence is hyperfine coupling with background nuclear spins (Figure 9a).[351] Isotopic purification has been experimentally shown to dramatically extend coherence lifetimes by orders of magnitude in other material systems such as silicon and carbon.[385–389] Theoretical works have also emphasized that zero nuclear spins are more critical in 2D systems compared to 3D systems. A 16-time improvement in coherence lifetime is predicted for isotopically purified monolayer $MoS_2$ with 2.76% net nuclear spins compared to its natural abundance con- centration (Figure 9b).[390] The nuclear spins of both the active 2D semiconductor and adjacent dielectric lay-ers can also be equally important. In 2D $WS_2$, qubit dynamics are predicted to be completely determined by adjacent dielectric nuclear bath (Figure 9c).[391]

### 1. Amorphous oxides

Amorphous oxide dielectrics were used in early gate defined TMD quantum dot devices (Figure 9d-f).[392, 393] ALD grown $Al_2O_3$ with thicknesses of 40 nm and 100 nm were used to realize gate defined quantum dots in $WS_2$ and $WSe_2$. Coulomb blockade was observed indicating single-electron tunnelling and carrier confinement. However, these early devices were made on thick quasi-2D TMD exfoliated flakes (>7 layers), due to cryogenic contact and dielectric fabrication challenges with 2D TMDs (1-3 layers). Independent gate tunability, control down to the last electron, and excited state measurements were not achieved with these thick $Al_2O_3$ More recently, 20 nm ALD $HfO_2$ was used to realize gate defined quantum dot in CVD grown bilayer $WS_2$.[90] Notably, this work reports gate defined quantum dots using non-exfoliated materials. However, the disadvantage of ALD grown oxides is still evident from the cross-sectional transmission electron micrographs showing rough interfaces. Low temperature measurements also confirm interface roughness limited transport. While some degree of independent gate control was demonstrated, the Coulomb blockade measurements showed irregular edges indicative of disorder and unintentional dot formation.

### 2. Crystalline dielectrics

Device performance is frequently improved by encapsulating 2D TMD flakes with crystalline hBN dielectrics(Figure 9g-i). Encapsulation with hBN enabled independent gate control, offering tunability over tunnelling barriers and quantum dot chemical potentials.[394–397] With improved gate control, double quantum dots were also demonstrated through measurements of characteristic hexagonal honeycomb stability diagrams.[394, 395] However, while few-carrier regime is still not reached, best efforts of an estimated 10-20 carriers was still sufficiently low to observe excited states in transport.[396]

### 3. Outlook

QIP applications demand devices of the highest quality that can operate at millikelvin temperatures. Current state-of-the-art devices rely heavily on crystalline hBN, but the lack of stable zero nuclear spin isotopes for boron and nitrogen is a fundamental limitation. This also rules out fluoride-based dielectrics such as CaF$_2$. Furthermore, the large effective mass of carriers in TMDs demands increasingly smaller lithographic device geometries to achieve effective carrier confinement. Dielectrics that offer small EOTs and the possibility to fabricate overlapping gate structures with minimal gate leakage currents will be useful. In this regard, the small dielectric constant of hBN is also a major disadvantage.

Insulating metal oxides can be ideal candidates. Several high-$\kappa$ variants exist while oxygen has a naturally high abundance of zero nuclear spin isotopes (> 99.9%). High-$\kappa$ oxide candidates that offer the potential for zero nuclear spin isotopes include HfO$_2$, ZrO$_2$, TiO$_2$, and MoO$_3$. Perovskites like SrTiO$_3$ and silicates like HfSiO$_4$ and ZrSiO$_4$ may also be compatible. The main challenge will be the integration of dielectric/TMD heterostructures with clean and smooth interfaces. Here, liquid metal synthesized oxides may offer a solution. Several crystalline oxides including HfO$_2$ have already been experimentally demonstrated.[70, 71] Atomically smooth interfaces should also be possible by leveraging the conformal nature of liquid metals to directly print the oxides over 2D TMDs.

### H. Quantum Sensing

Quantum sensing refers to the measurement of a physical quantity (classical or quantum) using a quantum system or a quantum property of the sensor.[398] Deploying 2D materials in sensing experiments is advantageous when high surface-to-bulk ratio,[399, 400] close proximity to the sensing targets,[401] or properties of the 2D material [402] are desirable for the application. We can broadly classify quantum sensing experiments with TMDs based on the degrees of freedom leveraged. First, precise control and measurement of the location and flow of electrons in TMDs (see previous section on Quantum Information Processing) allows the use of single electrons confined in gate-defined quantum dots as sensors, e.g., charge sensing sensitive down to the single electron level. Second, localized quantum defects in TMDs have optically active transitions sensitive to their environment.[403–405] This presents opportunities for using the defects as optically addressable quantum sensors, similar to what have been demonstrated with the nitrogen-vacancy centers in diamond.[406–409]

In this section, we will discuss the roles of dielectrics for 2D TMD based quantum sensors. Although there may be application-specific requirements for the dielectrics, the general desirable properties of the dielectrics can be summarized as: (i) negligible perturbations to the initialization, manipulation, and readout of the quantum sensor, (ii) minimal additional noise to the system. We aim to identify the optimal properties of dielectrics for each application.

### 1. Nanoelectronic quantum sensors

2D TMD-based nanoelectronic devices can be deployed as nanoscale magnetometers or charge sensors. For magnetic field sensing, gate-defined quantum dots can be used where the sensing target is approximated as a dipole and its magnetic field decreases with distance $r$ from the sensor as $\sim 1/r^3$.[410, 411] Therefore, it is beneficial to position the magnetometer close to the sensing target. For gate-defined quantum dots made using bulk semiconductor heterostructures such as GaAs/AlGaAs or Si/SiGe, the two-dimensional electron gas lies 50 - 100 nm below the surface (Figure 10a). In comparison, the TMD quantum dot can be as close as 20 nm away from the top surface,[90] limited by the thickness of gate dielectric. In single gate architecture, it may be possible to heterogeneously integrate the TMD with the sensing target and achieve sub-nm distance between the sensor and the target for increased sensitivity.

This is similar for charge sensing which aims to measure the charge displacement in solid-state qubits for quantum information processing applications.[352] With spin-to-charge conversion, electron spin readout is also possible. The charge detector can be realized with a quantum point contact [412] or a quantum dot/ single electron transistor (Figure 10b).[413, 414] Although the field of TMD quantum dots is still in its infancy, expected research trends may be similar to that of bulk semiconductors where the development and integration of charge sensors eventually led to high-fidelity, single-shot measurements of the charge and spin states.[415] The requirements for millikelvin operations and low-noise wiring preclude this type of quantum sensor from biological and environmental sensing applications.[416] However, it can be used as a sensitive local probe for emerging physical phenomena in condensed matter systems, such as Kondo effect [417], quantum Hall edge state [418], and topologically-protected Majorana zero modes [419].

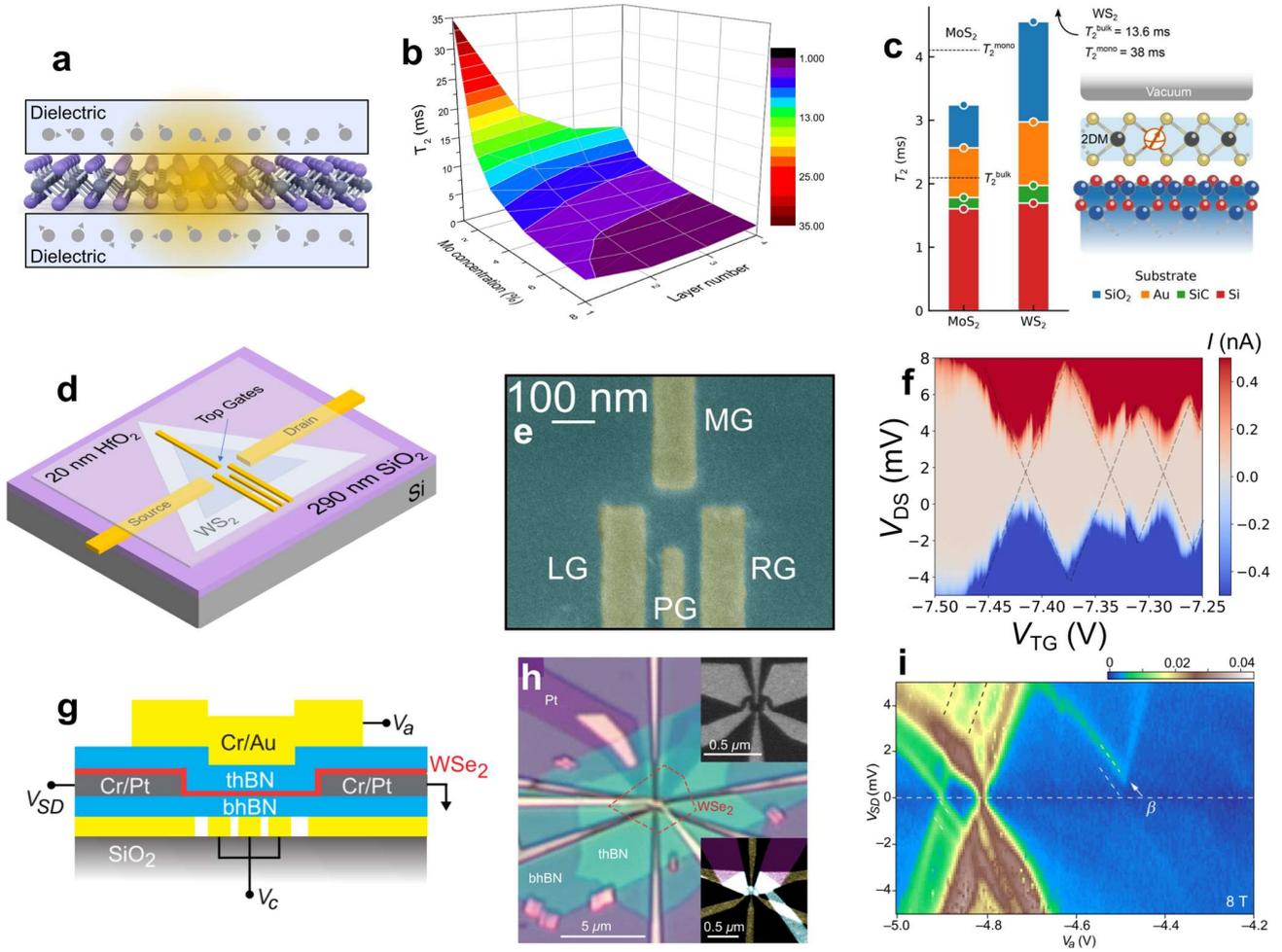

FIG. 9. (a) Schematic of nuclear spin coupling from the substrate to the 2D TMD layer. (b) Calculated coherence lifetimes $T_2$ for multi-layer MoS$_2$ as a function of Mo nuclear spin concentrations and number of layers, showing enhanced $T_2$ with lower nuclear spin concentrations; this enhancement is more effective with fewer layers. (c) Calculated $T_2$ for MoS$_2$ and WS$_2$ on different substrates. $T_2$ values are improved for monolayers compared to bulk. Longest (shortest) $T_2$ values are obtained for SiO$_2$(Si) which has the lowest(highest) concentration of $^{29}$Si nuclear spins. The small gyromagnetic ratio and large nuclei separation of Au results in moderate substrate influence of $T_2$ despite the highest concentration of nuclear spins. SiC presents an additional source of nuclear spins from $^{13}$C. In WS$_2$, the qubit dynamics is dominated by the substrate nuclear bath. Schematic of gate-defined 2D TMD based quantum dots using (d) amorphous HfO$_2$. (e) Scanning electron micrograph of gate geometry. (f) Diamond shaped patterns in the current stability diagram are indicative of single-electron transport due to Coulomb blockade. (g) A single quantum dot encapsulated with crystalline hBN. (h) Optical image of the device, the insets show the scanning electron micrograph of the gate geometry. (i) Excited states from Zeeman splitting with an applied magnetic field of 8 T can be observed. (b) Reprinted with permission under a Creative Commons Attribution 4.0 International License from ref[390]. Copyright 2019 Springer Nature. (c) Reprinted with permission from ref[391]. Copyright 2021 AIP Publishing. (d-f) Reprinted with permission from ref [90]. Copyright 2021 John Wiley and Sons. (g-i) Reprinted with permission from ref [396]. Copyright 2020 American Physical Society.

For TMD charge detectors or magnetometers, dielectric requirements are similar to those of TMD-based QIP (see Quantum Information Processing section), namely nuclear spin-free and high-$\kappa$ dielectrics. However, more efforts may be needed to understand the effects, if any, of thinner dielectrics on the relaxation and coherence times of TMD spin qubits. To date, there are few works that study the impact of noise from top dielectric surfaces further away from the TMD quantum dots. As we move towards thinner dielectrics for quantum sensing applications, it is possible that charge noises and magnetic noises from the dielectric top surface influence the spin relaxation and coherence times of the quantum dots, similar to what have been observed in other quantum sensors.[420–422] As the magnetic field sensitivity of a quantum sensor also depends on the spin relaxation or coherence time [398], adverse effects on the top surface of dielectrics can negate the benefits from having thinner dielectrics.

### 2. Optically addressable quantum sensors

Optically addressable quantum sensors refer to the group of sensors where the spin degrees of freedom are coupled to optical transitions. For solid-state sensors, this is usually realized with localized, point-like defects. The main advantages of optically addressable sensors are high spatial resolution and relatively relaxed operating requirements. [416] As demonstrated with nitrogen-vacancy centers in diamond, such atomic-scale sensors could be used as sensitive magnetometers (Figure 10c) to probe condensed matter systems [428], biological systems,[429, 430] and geological samples [431]. Recent reports have shown that TMDs can also support spatially confined excitons (Figure 10d).[403–405, 432] Although the spin-photon interface still needs to be established, this raises interesting possibilities of harnessing quantum defects in TMDs as optically addressable quantum sensors.[402] Future works in this direction are expected to follow that of defects in hBN, where the spin-photon interface was recently demonstrated (Figure 10e).[427, 433]

Optically addressable TMD quantum sensors could offer several advantages over existing solid-state atomic sensors. Compared to nitrogen-vacancy centers, the 2D nature of TMDs allows more efficient photonic device integration that could lead to higher sensor readout fidelity. [434] Compared to defects in hBN, TMD quantum sensors can have longer spin coherence times by benefitting from isotopic purification. In addition, TMDs exhibit valley degrees of freedom that couples to the polarization of emitted photons, and could be exploited in optical sensing protocols.

In principle, the operation of optically addressable quantum sensors does not require dielectrics. However, previous work has demonstrated enhanced device quality with full crystalline hBN encapsulation. Similar encapsulation with nuclear spin-free dielectrics could potentially enhance spin coherence times and in turn the sensitivity of TMD quantum sensors. An additional requirement is that the dielectrics must be optically transparent at the emission wavelengths of the sensors. On the other hand, high-$\kappa$ dielectric requirement could potentially be relaxed for optically addressable TMD quantum sensors that do not require gate operations.

### 3. Outlook

Although research on TMD-based quantum sensors is currently intensifying, more studies are required to establish the efficacy of these sensors. In the quest for optimal dielectrics that meet the requirements for quantum sensing applications, efforts should be made along two routes: demonstrations of quantum sensing applications with TMD-based sensors using crystalline hBN dielectrics, and exploration of alternative nuclear spin-free dielectrics.

## I. Valleytronics & Spintronics

While electronics utilize an electron's charge to encode bits of information, spintronics exploits the electron spin degree of freedom.[435, 436] Valleytronics employs the valley degree of freedom, i.e., a state associated with the local extrema in the electronic band structure.[437] Valleytronics is an emergent field that has become a subject of intense research with the rise of layered 2D materials. In many 2D systems like graphene or monolayer TMDs, spin and valley degrees of freedom are strongly coupled.[438, 439] Hence, spin- and valleytronic devices often share similar designs and architectures and their technological advances often rely on similar material requirements. Spintronics and valleytronics target various technological concepts for processing and storing information.[440, 441] In particular, 2D spin- and valley-based device geometries provide intriguing opportunities for compact memory elements,[442, 443] 'Beyond CMOS' logic devices,[444–447] and quantum computing technologies [448–451]. While the lead role in these devices is taken by graphene or 2D semiconductors, the substrate dielectrics are often designed to reduce the charge and spin scattering rates that strongly impact the spin transport [436] and valley polarization mechanisms. [449] For example, the use of hBN as a protective top layer yielded record-breaking spin diffusion length in graphene, reaching 30.5 $\mu m$ at room temperature.[452] Since the early works on hBN/graphene devices showed improved mobility and gate control [18, 453], encapsulation in hBN dielectric became the standard technique for high-quality 2D devices. Fully hBN-encapsulated monolayer $MoS_2$ devices have been used to demonstrate pure valley currents in monolayer TMDs [454], previously impossible due to low device quality [455]. Spin- and valleytronic devices greatly benefit from improved electronic properties of 2D materials driven by dielectric engineering described in previous sections.

We identified three critical architectures where dielectric engineering plays a crucial role. The first section covers dielectric barriers for effective spin injection and detection in lateral spin-valve devices. The next section surveys the routes to break the inversion symmetry in centrosymmetric 2D materials to enable effective (spin-)valley transistors. Here, the dielectric substrate can ei-ther induce lateral symmetry breaking via staggering potential; or provide an effective insulator layer for strong out-of-plane electric field tuning. The third section focuses on magnetic insulators in magnetic proximity effects. As valleytronics and spintronics is an emerging field for 2D TMDs, several experimental examples will be made using graphene to illustrate the underlying concepts.

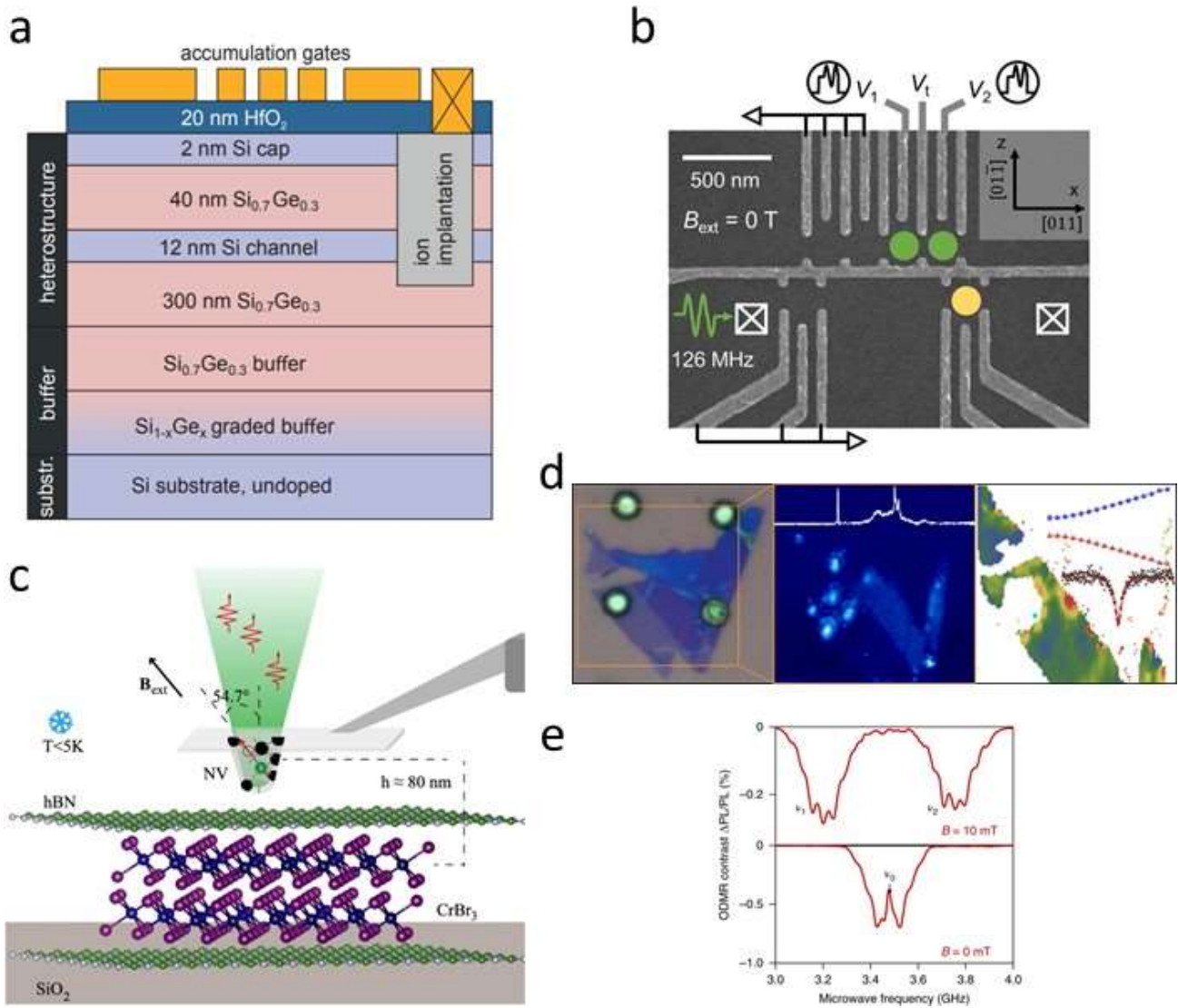

FIG. 10. (a) Cross-sectional schematic of bulk Si/SiGe heterostructures. The 2D electron gas is confined at the interface of Si channel, 62 nm below the top surface. For gate-defined quantum dot in TMDs, this could be reduced to less than 20 nm. (b) Scanning electron micrograph of gate-defined double quantum dots in GaAs (green) coupled to a charge sensor made by another quantum dot (yellow). This allows high-fidelity readout of the quantum states in the double quantum dots. (c) Schematic illustration of scanning magnetometry experiments with nitrogen-vacancy centers. The stray magnetic field of the $CrBr_3$ bilayer is measured using a single nitrogen-vacancy center in a diamond probe attached to the tuning fork of the atomic force microscope. The nitrogen-vacancy center can be brought 80 nm away from the sample. Optically addressable TMD sensors could further reduce the distance and increase the resolution in condensed matter sensing experiments. (d) Localized emission in $WSe_2$ created by defect-bound excitonic complexes. Second-order photon correlation statistics at zero-time delay, $g^{(2)}(\tau = 0)$, of 0.17 indicates the emission is from isolated, single quantum defect. The fine-structure splitting with an external magnetic field suggests high sensitivity to the environment. (e) Continuous-wave optically detected magnetic resonance spectrum for $V_{\bar{B}}$ centers in hBN, showing the spin-photon interface for quantum defects in 2D materials. (a) Reprinted with permission from ref[423]. Copyright 2019 American Chemical Society. (b) Reprinted with permission from ref[424]. Copyright 2021 American Chemical Society. (c) Reprinted with permission under a Creative Commons Attribution 4.0 International License from ref [425]. Copyright 2021 Springer Nature. (d) Reprinted with permission from ref [426]. Copyright 2015 American Chemical Society. (e) Reprinted with permission from ref [427]. Copyright 2020 Springer Nature.

### 1. Spin injection and detection

A typical 2D spin-valve device consists of a lateral conductive non-magnetic (N) channel with a set of ferromagnetic (FM) contacts (Figure 11a). Efficient spin injection and detection largely depend on the tunneling barriers that mitigate conductivity mismatch and spin absorption by FM contacts [469–471]. The spin injection polarization is expressed as $P = (R_F P_F + R_C P_C)/R_{tot}$, where $R_{tot} = R_F + R_N + R_C$ is total effective resistance, $R_F$, $R_N$ and $R_C$ are the resistances of FM electrode, N semiconductor, and contact, respectively. The $P_F = (\sigma_\uparrow - \sigma_\downarrow)/(\sigma_\uparrow + \sigma_\downarrow)$ and $P_C$, are the conductivity polarisation of the FM electrode and contact, respectively ($\sigma_\uparrow$ and $\sigma_\downarrow$ display the conductivity of spin-up and spin-down carriers). In the absence of tunneling barriers, $R_C = 0$, $P$ is low since $R_F \ll R_N$. For optimal spin injection $R_{C\geq} R_N$, which is satisfied for narrow tunneling barriers of atomic thickness. Early studies of spin injections into graphene have utilized amorphous metal oxide barriers, like $Al_2O_3$,[458] MgO,[472] $TiO_2$,[460] and SrO[473]. However, the difficulty to achieve thin homogeneous dielectric layers limited success. In contrast, crystalline dielectric tunneling barriers, like fluorinated graphene [461] or few-layer hBN [462] have shown superior injection. Bias-dependent FM/bilayer-hBN/graphene contacts have recently shown spin injection probabilities close to 100% [463], Figure 11b. Besides planar contact geometry, spin injection and detection in graphene can be achieved through one-dimensional edge contact thus allowing to effectively tune the contacts and the channel via the backgate [474].

While graphene offers outstanding spin transport properties like long spin coherence length [452], the lack of bandgap limits its use in semiconducting spin- and valleytronic devices. The zero-gap nature of graphene implies that the charge and spin currents cannot be fully suppressed, hampering the spin amplification and rectification mechanisms [475]. In contrast, 2D semiconductors like monolayer TMDs provide a broad set of materials with bandgaps ranging 1-3 eV and excellent gate tuneability and charge ambipolar characteristics [13, 476]. Compared to graphene, however, electrical injection of spins into TMDs so far shows limited progress. One distinctive difference between spin-valves based on graphene and monolayer TMDs is the effect of the tunneling barrier on the contact resistance. While the introduction of thin tunneling dielectrics in FM/graphene junction increases the $R_C$, in FM/TMD the resistance is reduced [477–480]. Moreover, broken inversion symmetry in monolayer TMDs inhibits the transport of in-plane spins due to an effective magnetic field that favors out-of-plane spin polarization [481, 482]. This is avoided in few-layered $MoS_2$, where the inversion symmetry is restored and a tunneling barrier like MgO allows effective spin injection [483].

### 2. Inversion symmetry breaking

A 2D valleytronic transistor is an (opto-)electronic logic device that processes the signal through the valley degree of freedom [484] with the possibility to operate at room temperature [485]. A key driving mechanism of the device relies on the broken inversion symmetry in 2D semiconductors.[486] Atomically thin materials like graphene, however, possess an intrinsic inversion symmetry due to their honeycomb lattices with two equivalent triangular carbon sublattices that can require additional engineering complexity to break. For example, this equivalence can be broken by a substrate that imposes a potential energy difference between two sublattices. Early works on epitaxial graphene have shown that sublattice symmetry can be broken by interlayer interaction between graphene and SiC substrate [487]. Closely matching lattice constants of 2D material and crystalline substrate supports the formation of a superlattice that ultimately creates a material with tailored electronic properties.[488] Similarly, monolayer graphene aligned with its isomorph, hBN ($\sim$1.8% lattice mismatch), experiences broken inversion symmetry [489–492] leading to the observation of valley-contrasting topological currents [493]. The amplitude of the inversion symmetry breaking can be controlled through the interlayer interaction strength with the substrate.[489, 494] Besides SiC and hBN, crystalline $Al_2O_3$ [495], diamond [496], $C_3N_4$ [497], black phosphorous [498], and $MoS_2$ [496] has been theoretically proposed as a substrate creating a commensurate superlattice with a large onsite potential difference and thus breaking the inversion symmetry in graphene.

The polar nature of many few-layered 2D materials, e.g., bilayer TMDs and AB-stacked graphene, allows symmetry breaking in centrosymmetric systems via a vertical electric field.[499] The vertical electric field is an effective knob to apply and tune an asymmetric interlayer potential across the layers, effectively breaking the inversion symmetry [500–502]. As a common device architecture, dual-gated 2D FETs allow independent tuning of the carrier density and electric field in the semiconducting channel (Figure 11c). In this geometry, the dielectric layers are expected to exhibit large breakdown voltages (in the order of V/nm) as well as minimal leakage currents. The difficulty of integrating conventional dielectrics with 2D materials imposes an engineering challenge. Through early studies, several dielectric layers have been tried in dual-gated geometry, including $SiO_2$[501, 503], $Al_2O_3$[504, 505], as well as vacuum in suspended devices[506, 507]. The vacuum barrier is particularly interesting as it provides a convenient way to clean the surfaces of the channel via current annealing. However, with the rise of vdW heterostructures [508], hBN became one of the most popular dielectric layers [509] that offers an atomically smooth interface with 2D materials and considerably high dielectric breakdown of $\geq 1V/nm$. [510] Encapsulated in hBN, dual-gated FETs of bilayer graphene show field-tunable bandgap - a manifestation of the broken inversion symmetry (Figure 11d), thus generating pure valley currents [466, 467, 511, 512]. In bilayer $MoS_2$, the vertical electric field induced valley polarization through the controllable breaking of the inversion symmetry [502, 513]. Full control over the electric field in bilayer systems thus enables one to amplify or switch valley currents on demand, an important step towards valleytronic logic devices.

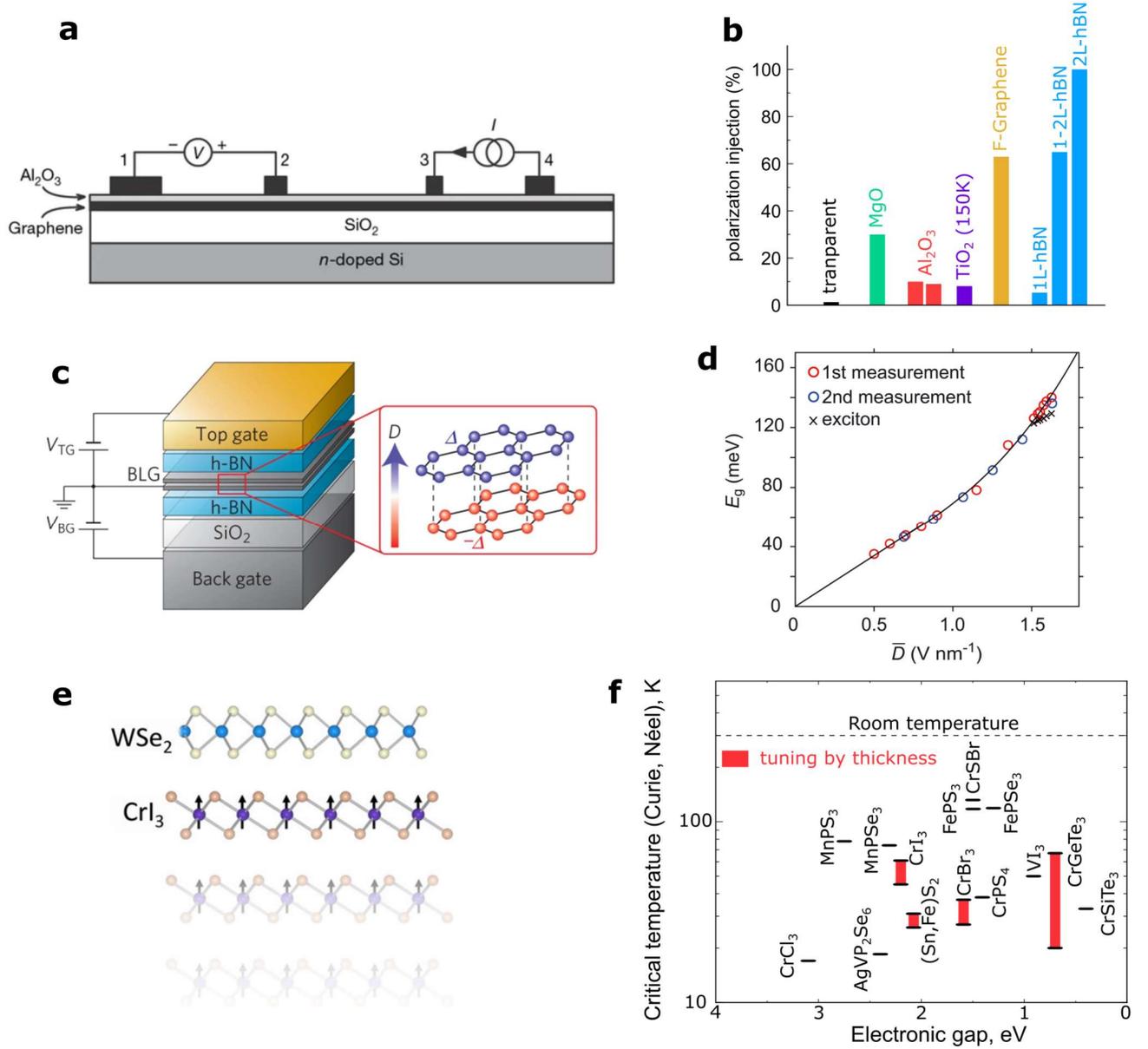

FIG. 11. (a) A lateral (non-local) spin valve device based on monolayer graphene. The current *I* is injected from electrode 3 through the Al₂O₃ barrier into graphene and is extracted at contact 4; (b) variation of spin polarization with the tunneling barrier material (The data are taken from refs.[456–464]); (c) (left) schematic of the dual-gated bilayer graphene device. The two hBN layers and a SiO₂ film are used as gate insulators. Using the top and bottom gates, the carrier density and vertical electric field are independently varied. (right) The lattice structure of AB-stacked bilayer graphene, where the vertical electric field breaks the inversion symmetry and the bandgap emerges; (d) the bandgap of bilayer graphene as a function of the vertical electric field; (e) a ball-and-stick model of all-2D vdW heterostructure of monolayer WSe₂ and magnetic CrI₃; (f) critical temperatures of vdW magnetic materials plotted against their electronic bandgap. The filled bar represents the change of critical temperature with the thickness (data taken from ref.[465]); (a) Reprinted with permission from ref.[458]. Copyright 2007 Nature Publishing Group; (c) Reprinted with permission from ref.[466] Copyright 2015 Springer Nature; (d) Reprinted with permission from ref.[467]. Copyright 2022 American Association for the Advancement of Science; (e) Reprinted with permission from ref.[468]. Copyright 2018 American Chemical Society.

### 3. Magnetic proximity effects

Strong valley polarization in 2D materials is key to realizing practical valleytronic devices.[437] While transient valley contrast can be achieved via optical excitations[514] or electrical spin injections,[515, 516] nonvolatile valley polarization requires an external magnetic field that lifts the valley degeneracy[437, 451]. However, valley splitting in monolayer TMDs requires high magnetic fields on the order of several teslas,[517–520] hampering the practical applications.

Alternatively, valley degeneracy can be lifted when a 2D TMD is deposited on a ferromagnetic substrate. Here, proximity-induced magnetic exchange interaction acts as an effective Zeeman field. Theoretical predictions estimate an effective field of thousands of teslas exceeding the strengths of state-of-the-art laboratory magnets.[521] Moreover, magnetic proximity effects potentially enables engineering ferromagnetic properties in a (typically) diamagnetic semiconductor,[522] thus realizing a magnetic semiconductor, a long-sought concept in spintronics.[523]

Initial experiments with graphene on bulk magnetic insulating substrates like yttrium iron garnet (YIG) or EuS revealed the proximity effects through the manifestation of the anomalous Hall effect [524] and Zeeman Spin Hall effect with exchange fields reaching $\sim 14$ tesla[525]. The weak overall magnetic exchange coupling has been explained by the low quality of the graphene/substrate interfaces.[526] Due to the short-range nature of magnetic exchange interactions, proximity-induced effects are strongly affected by the surface quality of the magnetic substrate. The recent discovery of the layered vdW magnetic materials [527–529] thus prompted interest in all-2D magnetic proximity devices. Indeed, the ideal vdW coupling in a heterostructure of a 2D semiconducting TMD with the 2D magnet can potentially offer an ultimate solution to the interface problem. [530, 531] Initial experiments have demonstrated electrical and thermal spin-currents in a bilayer graphene/CrSBr system with an effective Zeeman field $\sim 170$ tesla.[532] Since CrSBr is an interlayer antiferromagnet, individual FM layers of CrSBr couple antiferromagnetically to each other in the stack. As magnetic proximity coupling is largely driven by the interfacing layer, graphene experiences ferromagnetic proximity coupling. When monolayer TMD like $WSe_2$ is deposited on top of the vdW magnet $CrI_3$, magnetic exchange coupling induces valley splitting with effective Zeeman field of 13 tesla at a temperature of 5 K. [533] Akin to magnetic coupling in graphene/CrSBr, $WSe_2/CrI_3$ shows layer-resolved proximity interactions where the topmost layer dictates the coupling properties (Figure 11e), although the interlayer anti-FM structure of bulk $CrI_3$ cannot be completely ignored. [534]

### 4. Outlook

Despite the recent success of hBN in breaking the inversion symmetry, considerable leakage currents and low dielectric constant limit its use for practical implementations [139]. Moreover, the lack of lateral scalability of high-quality hBN narrows its application to proof-of-concept devices. Further studies are needed to identify a dielectric material with small equivalent-oxide-thickness and minimum interface states, the requirements for effective electric field control.[7, 10]

Besides the magnetic proximity functionality, vdW magnetic materials gained interest as an active material in magnetic tunnel junctions and spin-filters.[535] For example, $CrI_3$-based magnetic tunnel junctions exhibit tunneling magnetoresistance as high as nearly one million percent, [536] surpassing the giant magnetoresistance (GMR) of conventional multilayer structures by few orders of magnitude. [537] While vdW magnets *en masse* exhibit better coupling with 2D semiconductors than bulk magnetic insulators, their low critical temperatures (Curie or Neêl) limit practical applications (Figure 11f). However, exploration of the large family of vdW magnets has only just begun, with potentially nearly a hundred material systems waiting to be discovered.[538, 539] Moreover, vdW magnet properties can be tuned via twisting [531] or vertical electric fields [465], thus contributing additional control knobs in tailoring 2D materials for spin- and valleytronic devices.

## V. CONCLUSION

Expectations for 2D TMD applications in next-generation devices remain high, despite no commercial technology after over a decade of intensive research. However, similar to the historical development of silicon and its success over other semiconducting channel materials, it is likely that a major breakthrough will depend on the development of complementary dielectrics. It is important that research efforts are not focused solely on the search for alternative dielectric materials, but also con-sider the holistic requirements of their integration with 2D TMDs. Indeed, there are many existing well-known dielectrics that may be compatible with 2D TMDs, but integration techniques for these dielectrics need to be developed.

Research efforts should also reflect such considerations. Characterization and assessment of dielectric properties should be performed on 2D TMD/dielectric heterostructures instead of on only the dielectric. Stronger distinction can be made between mono-/bi- layer TMD devices with multi-layered ones, as device performance and dielectric integration compatibility can be dependent on the number of layers. More fundamental studies are needed to understand the physical mechanisms behind 2D TMD device degradation from dielectric integration, e.g., defect density and energy alignment. Several emerging dielectric materials have been identified in recent years, but their compatibility for large-scale integration with 2D TMDs remains unknown. Nevertheless, we remain optimistic that increasing research efforts into dielectrics for 2D TMDs will unlock their commercial potential.

## VI. VOCABULARY

Dielectric - A material or substance that is electrically insulating and can be polarized by applied electric fields. The ability of a dielectric to store electrical energy is quantified by its dielectric constant or relative permittivity.

Amorphous - A state or structure of a material lacking long-range order or crystalline pattern. In such materials, the atoms or molecules are arranged in random, disordered manner.

Exciton - A quasiparticle bound state of a negatively charged electron and a positively charged hole, i.e., a vacant electron state, in a semiconductor. An exciton forms when the electron in a material absorbs a photon and is excited to a higher energy state, leaving behind a positively charged hole.

Young's modulus - Also known as the elastic modulus or the modulus of elasticity, it is a mechanical property describing the stiffness/rigidity of materials and quantifies the relationship between stress and strain when a material is subjected to an applied force.

Quantum dot - Semiconductor particles with sizes on the order of manometers, which can have optical and electronic properties different from the bulk due to quantum confinement effects.

Biorecognition - The specific recognition and interaction between biological entities, resulting in the generation of a detectable signal.

Valley - The degree of freedom associated with the energy bands of materials with multiple local minima or maxima. They are distinct momentum states that electrons/holes can occupy.

## VII. GLOSSARY

2D - Two-dimensional
TMD - Transition metal dichalcogenide
CMOS - Complementary metal-oxide semiconductor
vdW - van der Waals
CVT - Chemical vapour transport
CVD - Chemical vapour deposition
hBN - Hexagonal boron nitride
ALD - Atomic layer deposition
MBE - Molecular beam epitaxy
LM - Liquid metal
FET - Field-effect transistor
EOT - Equivalent oxide thickness
$D_{IT}$ - Interface trap density
$D_{OT}$ - Border trap density
SS - Subthreshold swing
BTI - Bias temperature instability
FEOL - Front-end-of-line
BEOL - Back-end-of-line
NC-FET - Negative capacitance field-effect transistor
Fe-FET - Ferroelectric field-effect transistor
DRAM - Dynamic random access memory
BE - Binding energy
FoM - Figure of merit
EQE - External quantum efficiency
IQE - Internal quantum efficiency
LDR - Linear dynamic range
SNR - Signal-to-noise
LED - Light emitting diode
PI - Polyimide
PET - Polyethylene terephthalate
Bio-FETs - Bio field-effect transistors
PoC - Point of care
LoD - Limit of detection
QIP - Quantum information processing
T2 - Quantum decoherence dephasing time
NV - Nitrogen vacancy
FM - Ferromagnetic

This research was supported by the Agency for Science, Technology, and Research (A*STAR) under its MTC YIRG grant No. M21K3c0124. We acknowledge the funding support from Agency for Science, Technology and Research (#21709). I.V. and K.E.J.G. acknowledges support from a Singapore National Research Foundation Grant (CRP21-2018-0001). D.H. acknowledges funding support from A*STAR Project C222812022 and MTC YIRG M22K3c0105.